\documentclass[lettersize,journal]{IEEEtran}
\usepackage [latin1]{inputenc}
\usepackage{cite}
\usepackage[english]{babel}
\usepackage{epsfig}
\usepackage{algorithm}
\usepackage{algorithmic}
\usepackage{amsmath}
\usepackage{amsthm}
\usepackage{amssymb}
\usepackage{dsfont}
\usepackage{xcolor}
\usepackage{subfigure}
\usepackage{balance}
\usepackage{pdflscape}
\usepackage{placeins}
\usepackage{graphicx}
\usepackage{cases}
\usepackage{amsfonts}
\usepackage{multirow}
\usepackage{multicol}
\usepackage{mathrsfs}
\usepackage{color}
\usepackage{mathtools}
\usepackage{epstopdf}
\usepackage{algorithm}
\usepackage{float}
\usepackage{flushend}

\def\BibTeX{{\rm B\kern-.05em{\sc i\kern-.025em b}\kern-.08em
		T\kern-.1667em\lower.7ex\hbox{E}\kern-.125emX}}
\floatname{algorithm}{Algorithm}
\makeatletter
\begin{document}
	\title{System-Level Gains of Dual-Band RIS:\\ From Wireless Communication to SWIPT}
	\author{Aritra Basu, Zina~Mohamed,~\IEEEmembership{Member,~IEEE} and Sonia A\"{i}ssa,~\IEEEmembership{Fellow,~IEEE}
		\thanks{The authors are with the  Institut National de la Recherche Scientifique (INRS), Montreal, QC H5A 1K6, Canada; e-mail: \{aritra.basu, zina.mohamed, sonia.aissa\}@inrs.ca.}
		\thanks{This work was supported by a Discovery Grant from the Natural Sciences and Engineering Research Council (NSERC) of Canada.}
        \thanks{This work has been accepted for publication in IEEE Transactions on Communications. 2026 IEEE. Personal use of this material is permitted. Permission from IEEE must be obtained for all other uses.}
	}
	\maketitle
	\begin{abstract}
		This paper introduces the integration of dual-band reconfigurable intelligent surfaces (DBRIS) in wireless system design, and investigates the ensuing performance gains in different application scenarios. This type of metasurfaces offers a compelling technology due to their reduced material usage, compact design, and multi-functional operation across multiple frequency bands compared to single-band RISs. Two representative use cases are considered to assess the achievable performance gains with the said integration, namely, DBRIS-assisted wireless communication, and DBRIS-assisted simultaneous wireless information and power transfer (SWIPT). For the first design, where the DBRIS is deployed as a relay, we focus on the energy efficiency (EE) maximization of the communication system. An alternating manifold method is proposed for the joint optimization of the source transmit power and the DBRIS phase shifting. For the second design, we investigate a DBRIS-assisted SWIPT system using a novel power splitting approach. An EE metric is defined, and an alternating optimization algorithm is developed to maximize it. Numerical results validate the advantages of the proposed design architectures as well as the proposed optimization framework. In particular, we highlight the energy efficiency gains enabled by the DBRIS as compared to operations with conventional single-frequency surfaces, in both application scenarios.
	\end{abstract}
	\section{Introduction}
	\subsection{Context and Motivation}
	As the wireless ecosystem evolves towards the sixth generation (6G), ensuring high energy efficiencies while minimizing material consumption and promoting hardware sharing are becoming critical imperatives for the design and development of wireless systems and networks that are sustainable in and of themselves. These imperatives are driven by the rising footprint expected from the exponential growth of wireless connectivity demands in a plethora of applications such as autonomous systems, massive machine-type communications, extended reality, etc. \cite{9706187}.

To meet the service requirements without expanding the physical infrastructure of the underlying networks, intelligent metasurfaces, and particularly reconfigurable intelligent surfaces (RISs), have emerged as a key enabling technology, thanks to their passive operation, low hardware cost, ease of deployment, and high efficacy in non line-of-sight (LoS) propagation conditions \cite{vetu2910, lopp234}. Indeed, by leveraging arrays of passive elements governed by integrated circuits, RIS can dynamically manipulate the incident electromagnetic waves to transform the propagation environment from stochastic and unmanageable to a more deterministic and programmable medium \cite{bxzb}.

From an RF engineering standpoint, RIS elements can control the phase, amplitude, and polarization of reflected signals via tunable components such as PIN diodes and varactors \cite{sdbtt959}. This software-controlled behavior allows RIS to enhance link reliability, extend radio coverage, and reduce interference, while significantly lowering energy consumption by avoiding active RF chains, making the RIS technology well aligned with the goals of sustainable wireless networks \cite{Rana2023}.
	
	Owing to these advantages, RIS has been investigated for a wide range of applications and different system set-ups, including index modulation \cite{acsqqwqw}, device-to-device communications \cite{Bhowal}, networks assisted by unmanned aerial vehicles (UAV) \cite{aissa2020}, vehicle-to-vehicle systems \cite{sdbsas}, simultaneous wireless information and power transfer (SWIPT) \cite{10718349}, and ultra-reliable low-latency communication \cite{camad24}. From a deployment standpoint, RISs can be used as passive reflectors and as programmable access points \cite{gssksystems}, as well as hybrid architectures that combine reflection and modulation \cite{db15d1b}.
	
	Despite their initial success, conventional RIS designs face limitations such as the half-space signal coverage and the limited support for multi-functional operations, which restrict their scalability and adaptability in next-generation networks. These issues are further compounded by hardware complexity and rigid frequency selectivity, all of which limit the ability of conventional RIS architectures to operate across multiple frequency bands; a critical requirement for supporting the diverse service demands of future multi-band multi-service wireless systems. As wireless systems evolve into more heterogeneous infrastructures and the spectrum resources become increasingly scarce, 
    it is imperative to move beyond conventional RIS architectures to achieve higher spectrum utilization and sustainable infrastructure deployment.
	
	In this context, researchers have explored different RIS variants, including simultaneous transmitting and reflecting RIS, active RIS, and dual-polarized RIS, which offer broader angular coverage and signal control compared to conventional RISs \cite{db15d1b, zbbzb, wanliIoT, 10780951, 9931168, 9906843, ljoijj, dpchen, 10685084}. Nonetheless, these structures struggle with hardware complexity, cross-polarization interference \cite{4525659} and, above all, are limited to operation within a single frequency band.

In stark contrast, dual-band RIS (DBRIS) architectures leverage frequency-domain isolation to enable simultaneous operation over different frequency bands. By supporting independent beamforming and signal shaping over distinct bands, the DBRIS technology can enhance spectral utilization 
and link robustness for diverse services expected in future 6G networks, while reducing the material use and cost as compared to operations with conventional single-band RISs \cite{10777622,8993760,f1}.
	\subsection{Physical Realization of DBRISs}
	Recent studies 
    demonstrated promising physical implementations of DBRISs \cite{f1, johriRF, jhgnh, 8993760, Zou21, 10777622}. For instance, the work in \cite{f1} advanced a super-cell structure consisting of circular and square microstrip patches, each resonating at a specific frequency, and employed two low-cost varactor diodes to achieve frequency-tunable reactance. A gap is maintained between the said patches to accommodate a varactor diode placed between two circular rings with a given spacing. The experiments demonstrated improved reflection performance with minimal cross-coupling owing to the frequency-separated excitation, and the design's potential in achieving high information-carrying capacity. The work in \cite{johriRF}, on the other hand, presented a low-loss miniaturized DBRIS unit cell ($13 \times 13~{\rm mm}^2$) for operations at 5.8 GHz and 7.3 GHz, which integrates a butterfly-stub RF choke to suppress radiation losses while maintaining $180^{\circ}$ phase modulation per surface element. For independent dual-band control, the authors in \cite{jhgnh} designed a 3-bit DBRIS with varactor-loaded nested patches that can independently adjust the reflection phases at 2.52 GHz and 3.4-3.5 GHz, achieving a loss of $\leq$ 2 dB and an angular stability for incidence angles up to $50^{\circ}$. Other advanced designs, such as the one proposed in \cite{8993760}, further enabled dual-band real-time beam steering through modular control circuits, supporting orthogonal polarizations for simultaneous multi-channel operations. More recently, a DBRIS design that combines functionalities over mmWave and sub-6GHz bands within a single aperture was proposed in \cite{10777622}.

 	A challenge in designing DBRISs is mitigating inter-band electromagnetic coupling, which can otherwise degrade isolation and limit the independent tunability of the surface at its operating frequencies. Such coupling arises due to interactions between resonant elements tuned for different frequency bands, which can result in signal distortion, reduced reflection efficiency, and impaired frequency selectivity. To address this issue, structures such as a metallic ground plane were proposed. In fact, it has been demonstrated that structures such as defective ground plane, electromagnetic band-gap (EBG), etc., can inhibit the effect of coupling, hence their wide use in the design of DBRISs \cite{Zou21, article, 10777622, app14072928}. Indeed, EBGs can provide high isolation between frequency-selective elements by acting as spatial filters, thereby ensuring independent control over each frequency band \cite{10777622}. Also, metallic back reflectors, such as ground planes or frequency-selective surfaces (FSS), can be employed beneath the RIS substrate to redirect the backward-propagating waves and reduce the inter-layer interactions \cite{app14072928}. Additionally, vertical shielding structures and multi-layered substrate configurations with intermediate dielectric or magnetic spacers can be utilized to physically separate the resonant paths of the dual-band elements, further enhancing the inter-band decoupling \cite{10777622}. Together, these techniques ensure that the dual-frequency functionalities of DBRISs can coexist with minimal interference, thus enabling reliable operation in practice.
	\subsection{System-level Integration of DBRISs}	
	Despite the aforementioned advances in the physical realization of DBRISs, their integration into wireless systems remains an uncharted territory. This research gap presents a timely opportunity to transition DBRISs from the laboratory into wireless system design and development. Indeed, grounded on its physical feasibility, the DBRIS technology holds immense potential to enhance the capabilities and sustainability of wireless systems and networks in different application domains.

Chief among the advantages of the DBRIS technology is that it enables relaying functionalities not only for multi-band services within the same wireless system but also a shared use between different systems operating with different frequencies, thereby improving efficiency in terms of material utilization and the physical footprint of the deployment, while enhancing the quality of the supported services. In fact, thanks to their reconfigurability property, DBRISs can be used to mitigate blockages between transmitters and receivers by establishing alternative LoS links and ensuring reliable signal propagation between them. Such capabilities make them particularly attractive for integration into multi-band systems as a relay or transmitter.

Thanks to their reconfigurability property, DBRISs can be used to mitigate blockages between transmitters and receivers by establishing alternative LoS links and ensuring reliable signal propagation between them. Integration opportunities for DBRIS as a relay can be classified based on whether it is controlled by a single operator or shared among multiple operators. An application example in the single-operator scenario is leveraging DBRIS to support cellular services over multiple bands, e.g., 4G and 5G, in dense outdoor environments. The DBRIS technology can also be integrated to enable applications in indoor environments. For instance, the surface structure in \cite{10777622} can be exploited for non-overlapping functionalities such as WiFi-based positioning and localization across sub-6 GHz and mmWave bands. The multi-operator scenarios, on the other hand, involve the sharing of DBRIS across different service providers operating over distinct frequency bands, e.g., a 4G band for Operator A, and a 5G band for Operator B. RF implementations suitable for such cases are discussed in \cite{f1}. In indoor environments, DBRIS can also be shared to improve the performance of coexisting systems such as WiFi and visible light communication (VLC), e.g., to simultaneously enhance the RF coverage by mitigating blockages and enhance the VLC link reliability.
\subsection{Contributions}
	Building upon the preceding discussion, and the merits of multi-band metasurfaces in offering a more sustainable and scalable alternative to conventional single-band RISs (SBRISs), including reduced hardware, energy and material costs,\footnote{According to GSMA, dual-band radio and antenna systems incur only ~50\% higher total cost of ownership compared to single-band setups, making multi-band integration a cost-efficient strategy.} this paper proposes the integration of DBRIS in wireless system design and investigates the achievable performance gains in two representative use cases, each involving a dual-band transmitter and two single-band receivers, with the service assisted via a shared DBRIS operating at two frequency bands. The first use case focuses on data communication to the information devices, and the second considers SWIPT to an information receiver and an energy harvesting device.

The proposed design framework begins with the derivation of the DBRIS frequency-dependent response function using a two-ray propagation model, which captures the role of the intelligent surface in shaping the multipath environment across distinct frequency bands. Then, taking energy efficiency (EE) as a performance metric, given its importance from a sustainability standpoint, we focus on its maximization in both use cases. \footnote{The principal advantage of the DBRIS technology does not lie in creating fundamentally new beamforming degrees of freedom compared to two independent SBRISs, but rather in achieving comparable beamforming capability with substantially lower energy and material cost. By replacing two physically separated RISs with a single metasurface, the core benefit of the DBRIS technology is thus inherently architectural, with simplified control and reduced overhead requirements compared to system operation with separate SBRISs. Consequently, the focus of this work is not on throughput quantification but on demonstrating the sustainability and EE advantages of the DBRIS architecture.} The optimization problems are formulated and solved using alternating optimization approaches, whereby the transmit power allocation and the DBRIS phase shifts are jointly optimized in both systems under study, in addition to the optimization of the power splitting ratio in the SWIPT system.
	
	In detailing these contributions, the next section delves into the DBRIS response function. The design and optimization of the DBRIS-aided communication system is then developed in section III, and the one pertaining to the DBRIS-assisted SWIPT is provided in section IV. The simulation results are discussed in section V, followed by the paper's conclusion.
	
	{\it Notations:} Vectors and matrices are denoted by boldface lower- and upper-case letters, respectively. The operators $(\cdot)^{\mathrm{T}}$ and $(\cdot)^{\mathrm{H}}$ represent the transpose and the conjugate transpose, respectively. The symbol $\lVert \cdot \rVert$ denotes the Euclidean norm of a vector, and $\lvert \cdot \rvert$ indicates the modulus of a complex number. The mathematical expectation is denoted by $\mathbb{E}[\cdot]$, the Riemannian gradient is represented by $\nabla_{\mathcal{M}}$, and $\mathrm{Retr}$ is used to denote the retraction operator.
	\section{DBRIS Response Function}
	We adopt the super-cell based DBRIS architecture proposed in \cite{f1, jhgnh}, in which each unit consists of two distinct asymmetric patches, each equipped with independent biasing lines for frequency-selective operation. The DBRIS is engineered to operate simultaneously at two sufficiently separated frequency bands, denoted by $f_{\rm l}$ and $f_{\rm h}$ for the low and high frequencies, respectively. The separation between these, specifically 2.5 GHz and 3.6 GHz, is significant, thereby ensuring negligible electromagnetic coupling and mutual interference between the bands.\footnote{To mitigate the impact of residual inter-band coupling, the DBRIS is assumed to incorporate coupling-inhibiting structures such as EBG materials and metallic reflectors \cite{Zou21}.} Furthermore, spatial separation of the receivers, combined with independent beamforming at each frequency, renders inter-band interference negligible.
	
\begin{figure}[htb]
\centering
\includegraphics[height=4.7cm, width=1\linewidth]{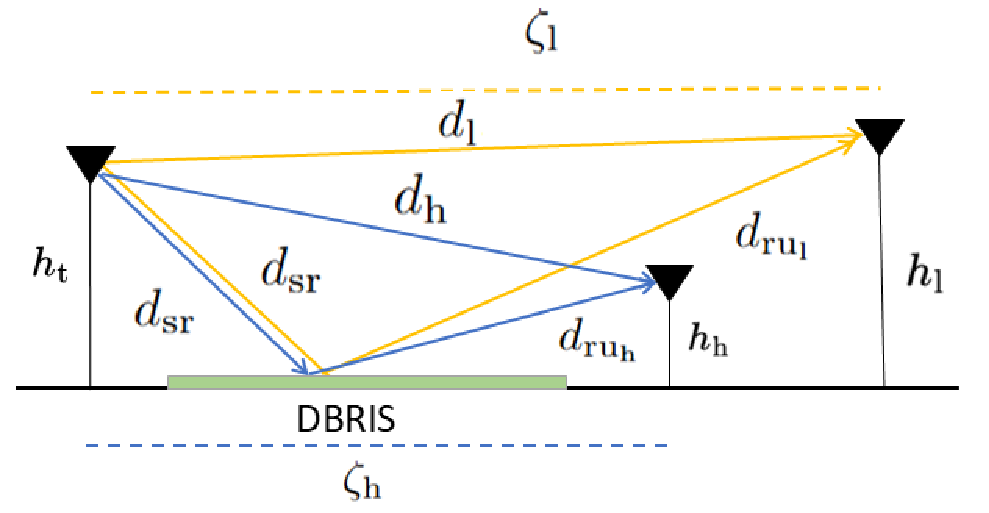}
\caption{Two-ray propagation geometry with a DBRIS acting as a specular planar reflector.}
\label{figa11}
\end{figure}
	
The response of the $i^{\rm th}$ DBRIS element at the lower operation frequency $f_{\rm l}$ is expressed as $\beta_{{\rm l},i} e^{-\jmath \phi_{{\rm l},i}}$, where $\beta_{{\rm l},i}$ is the amplitude reflection coefficient and $\phi_{{\rm l},i}$ is the tunable phase shift. At the higher operation frequency, $f_{\rm h}$, the element response is given by $\beta_{{\rm h},i} e^{-\jmath \phi_{{\rm h},i}}$, with $\beta_{{\rm h},i}$ and $\phi_{{\rm h},i}$ representing the amplitude and phase components, respectively. With a total of $N$ elements, these responses are collectively represented using the following diagonal matrices:
	\begin{align}
		\mathbf{\Phi}_{\rm l} &={\rm diag} \left(\beta_{{\rm l},1} e^{-\jmath \phi_{{\rm l},1}},\cdots,\beta_{{\rm l},N} e^{-\jmath \phi_{{\rm l},N}}\right),\\
		\mathbf{\Phi}_{\rm h} &= {\rm diag} \left(\beta_{{\rm h},1} e^{-\jmath \phi_{{\rm h},1}},\cdots,\beta_{{\rm h},N} e^{-\jmath \phi_{{\rm h},N}}\right).
	\end{align}
	Therefore, the overall DBRIS phase-shift matrix can be expressed as
	\begin{equation}
		\mathbf{\Phi} =
		\begin{bmatrix}
			\mathbf{\Phi}_{\rm l} & \mathbf{0} \\
			\mathbf{0} & \mathbf{\Phi}_{\rm h}
		\end{bmatrix}.
	\end{equation}
	For analytical tractability and given that our focus is on assessing the fundamental performance limits of the aforementioned DBRIS-aided wireless systems, we assume lossless DBRIS elements such that $|\beta| = |\beta_{{\rm l},i}|=|\beta_{{\rm h},i}|=1$.
	
	As illustrated in Fig. \ref{figa11}, with regard to frequency $f_{\rm l}$, the direct distance between the transmitter, referred to as the access point (AP), and receiver $u_{\rm l}$ is denoted by $d_{\rm l}$. The distances from the AP to the ground reflection point, and the one from the ground reflection point to the corresponding receiver, are represented by $d_{\rm sr}$ and $d_{{\rm ru_{\rm l}}}$, respectively. For the operation frequency $f_{_{\rm h}}$, the distance between the AP and receiver $u_{\rm h}$ is denoted by $d_{\rm h}$, and the distances of the ground reflected paths are indicated by $d_{\rm sr}$ and $d_{{\rm ru}_{\rm h}}$, respectively. The aerial distances between the AP and the receivers are indicated by $\zeta_{\rm l}$ and $\zeta_{\rm h}$ w.r.t. the frequencies $f_{\rm l}$ and $f_{_{\rm h}}$, respectively. In practice, the ground reflection coefficients, i.e., $|\beta_{{\rm l},i}|$ and $|\beta_{{\rm h},i}|$, are frequency-dependent parameters influenced by the material properties of the reflecting surface, the angle of incidence, and the frequency-selective absorption and attenuation characteristics of the electromagnetic waves \cite{9935151212}.
	
	We adopt a two-ray propagation model, in which the DBRIS is modeled as an electrically large planar reflector such that the geometric size of the DBRIS is larger than that of the radio waves' wavelength. In other words, we assume that the DBRIS is sufficiently large compared with the transmission wavelength, and it produces only specular reflections. As illustrated in Fig. \ref{figa11}, the received signal consists of a direct LoS component and a dominant specularly reflected component from the DBRIS surface, where the reflection obeys geometric optics and mirror-like propagation assumptions. Reflections are modeled based on the Snell's law of reflection.
	
Under narrowband transmission assumptions, the received signals at any observation time $t$ can be expressed as follows:
\begin{align}
		&y_{\rm h} \left(t\right) = \frac{\lambda_{_{\rm h}}}{4\pi}
		\left(\frac{e^-\frac{\jmath2\pi d_{\rm h}}{\lambda_{_{\rm h}}}}{d_{\rm h}} + \frac{\beta e^-\frac{\jmath2\pi \left(d_{\rm sr} + d_{\rm ru_{h}}\right)}{\lambda_{_{\rm h}}}}{d_{\rm sr} + d_{\rm ru_{h}}}\right) x_{\rm h}(t),\\
		&y_{\rm l} \left(t\right) = \frac{\lambda_{_{\rm l}}}{4\pi}
		\left(\frac{e^-\frac{\jmath2\pi d_{\rm l}}{\lambda_{_{\rm l}}}}{d_{\rm l}} + \frac{\beta e^-\frac{\jmath2\pi \left(d_{\rm sr} + d_{\rm ru_{l}}\right)}{\lambda_{_{\rm l}}}}{d_{\rm sr} + d_{\rm ru_{l}}}\right)x_{\rm l}(t),
	\end{align}
	\noindent where, for the frequencies $f_{\rm h}$ and $f_{\rm l}$, $x_{\rm l}$ and $x_{\rm h}$ are the baseband transmitted signals, $\lambda_{\rm h}$ and $\lambda_{\rm l}$ are the wavelengths, and where $\tau_{\rm l}$ and $\tau_{\rm h}$ are the relative propagation delays between the LoS path and the reflected path, expressed as
	\begin{align}
		\tau_{\rm h} = \frac{\left(d_{\rm sr} + d_{{\rm ru}_{\rm h}} - d_{\rm h}\right)}{c},~~\tau_{\rm l} = \frac{\left(d_{\rm sr} + d_{{\rm ru}_{\rm l}} - d_{\rm l}\right)}{c}.
	\end{align}
	\noindent in which $c$ denotes the speed of light.
	
	Consequently, the power of the signals at the two end receivers can be expressed as
	\begin{align}
		&P_{\rm u_h} = P_{\rm h}
		\left(\frac{\lambda_{\rm h}}{4 \pi}\right)^2
		\left|\frac{1}{d_{\rm h}} + \frac{\beta \delta_{\rm h}}{d_{\rm sr} + d_{{\rm ru}_{\rm h}}}\right|^2,\\
		&P_{\rm u_l} = P_{\rm l}
		\left(\frac{\lambda_{\rm l}}{4 \pi}\right)^2
		\left|\frac{1}{d_{\rm l}} + \frac{\beta \delta_{\rm l}}{d_{\rm sr} + d_{{\rm ru}_{\rm l}}}\right|^2,
	\end{align}
	\noindent where $P_{\rm h}$ and $P_{\rm l}$ are the transmit powers at $f_{\rm h}$ and $f_{\rm l}$, respectively, and where $\delta_{\rm h}$ and $\delta_{\rm l}$ are given by
	\begin{align}
\delta_{\rm h} =e^{-\frac{2\pi \left(d_{\rm sr} + d_{{\rm ru}_{\rm h}} - d_{\rm h}\right)}{\lambda_{\rm h}}},~~\delta_{\rm l} =e^{-\frac{2\pi \left(d_{\rm sr} + d_{{\rm ru}_{\rm l}} - d_{\rm l}\right)}{\lambda_{\rm l}}}.
	\end{align}
	
	Under far-field propagation conditions, we can assume that $\zeta_{\rm l} \approx d_{\rm l} \approx d_{\rm sr} + d_{{\rm ru}_{\rm l}}$ and $\zeta_{\rm h} \approx d_{\rm h} \approx d_{\rm sr} + d_{{\rm ru}_{\rm h}}$, due to the large distances involved. The $N$ elements of the DBRIS yield a gain of $(N+1)^2$ in the case where the LoS path exists, or $N^2$ otherwise. Accordingly, the signal powers at the end receivers can be expressed as follows:
	\begin{align}
P_{\rm u_h} \approx N^2 P_{\rm h}\left(\frac{\lambda_{\rm h}}{4 \pi \zeta_{\rm h}}\right)^2,
~~P_{\rm u_l} \approx N^2P_{\rm l}\left(\frac{\lambda_{\rm l}}{4 \pi \zeta_{\rm l}}\right)^2.
		\label{eq7a}
	\end{align}
	
The two-ray formulation highlights the multi-functional capability of the intelligent surface, namely, its ability to simultaneously reflect signals at distinct frequencies. However, optimal system performance critically depends on the appropriate allocation of the transmit powers $P_{\rm h}$ and $P_{\rm l}$. These power levels play a pivotal role in maximizing system performance metrics, such as throughput or EE, and must therefore be carefully optimized as part of the DBRIS-assisted wireless system design. While the two-ray deterministic formulation provides an idealized baseline that omits complex multi-path scatterers, terrain irregularities, and mobility, it yields essential physical insights into the hardware's core mechanics. To validate these insights under realistic deployment conditions like multi-path scatterers, etc., as mentioned earlier, the propagation channels in the subsequent optimization and simulation frameworks are modeled in strict accordance with the 3GPP standardized specifications \cite{3gpp38901}.
	\section{DBRIS Aided Communication}
	\subsection{System Model}
\begin{figure}[h]
		\centering
\hspace*{-0.05in}\includegraphics[height=2.5in, width=0.98\linewidth]{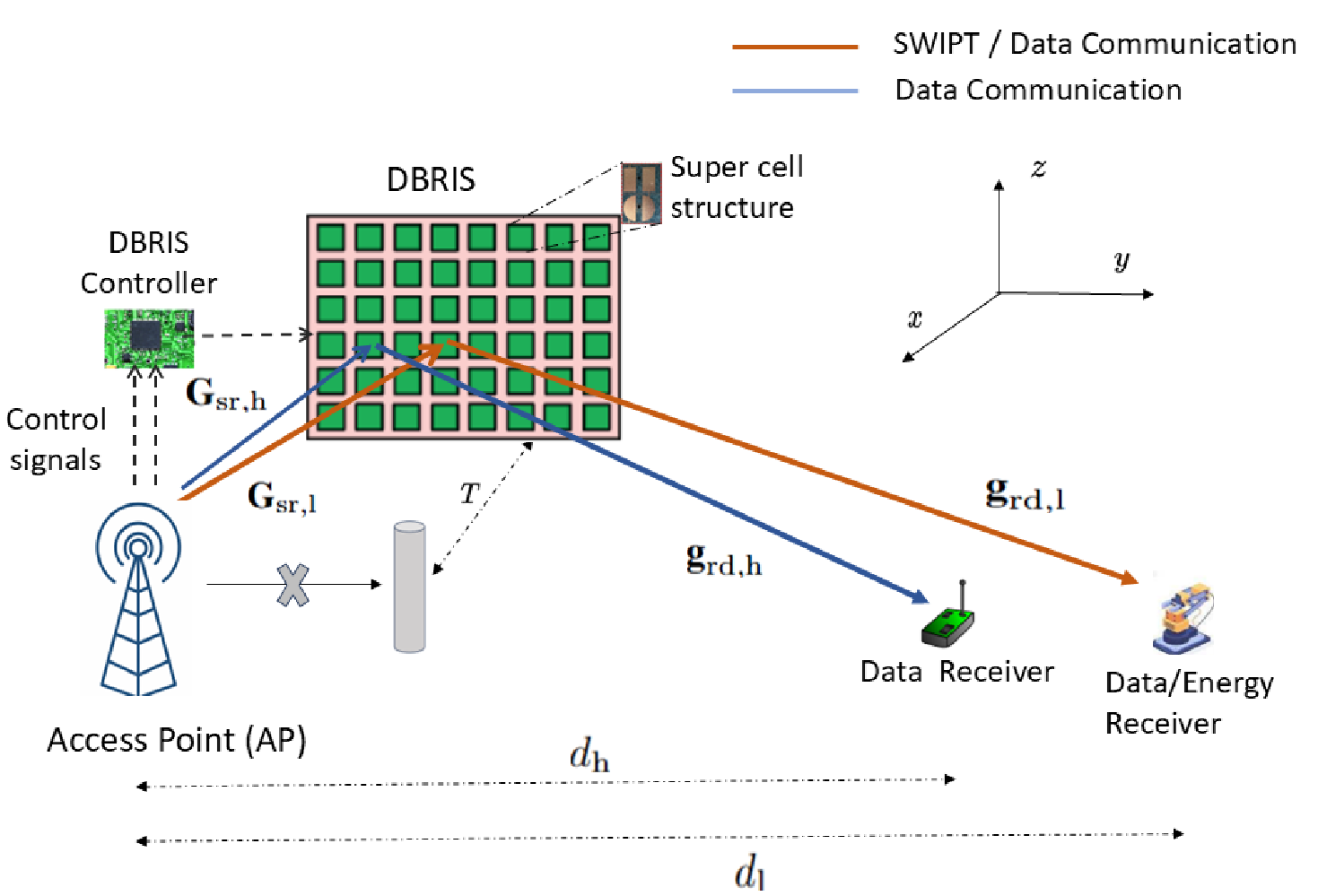}
		\caption{The DBRIS-aided system models. In the communication use case, the end receivers are both information devices. In the SWIPT model, the source transmits a data signal to an information device and wireless power to an energy harvesting device.}
		\label{figa1}
	\end{figure}

	The first use case under study is a DBRIS-aided downlink communication system comprising a common AP and two spatially separated narrowband receivers, as illustrated in Fig.~\ref{figa1}. The dual-band AP is equipped with two antennas, each dedicated to operation over a distant frequency band, and the end receivers are single-antenna devices. Assisted by the DBRIS, which is located between the AP and the receivers, the AP simultaneously emits two distinct signals over two distinct frequency bands, with negligible inter-band interference. The direct links between the dual-band AP and the end receivers are assumed to be either obstructed or significantly attenuated compared to the cascaded link via the DBRIS, allowing us to neglect the direct links without loss of generality. The receivers' positions are such that the far-field condition holds for the DBRIS.\footnote{Our focus being on the fundamental performance limits of the proposed DBRIS-assisted architecture, a two-user set-up is considered. Also, the AP's transceivers, users' receivers, and DBRIS, are considered ideal equipment. Extending the proposed framework to the multi-user scenario, where intra-band interference is taken into account, and considering equipment with hardware impairments is left for future work.}
	
The DBRIS, which is based on a super-cell architecture, is composed of $N$ elements, each integrating two varactor diodes controlled via independent bias voltages applied to separate pairs of metal patches \cite{jhgnh}, as discussed in section II. Further details about the hardware implementation of the DBRIS structure are provided in \cite{jhgnh}. So far, researchers proposed the use of ultra-narrowband metasurface antennas for multi-band RIS. With such a design, the antenna bandwidth can be reduced to less than 1\% of the carrier frequencies. Thus, isolation between the operation bands becomes viable  \cite{monemi2025ris}.

The super-cell based DBRIS is capable of simultaneously reflecting the two transmit signals towards their respective receivers. The DBRIS controller is connected to the AP with the help of a high-speed control link to transmit the channel information for passive beamforming at the DBRIS.

    For notational coherence, variables pertaining to the high and low frequencies, i.e., $f_{\rm h}$ and $f_{\rm l}$, are denoted by the subscripts ${\rm h}$ and ${\rm l}$, respectively. Denoting the transmitted signal in the frequency bands $f_{\rm h}$ and $f_{\rm l}$ by $x_{\rm h}$ and $x_{\rm l}$, respectively, the received signals can be expressed as
    \begin{align}
		y_{\rm h} &= \sqrt{P_{\rm h}}\textbf{g}_{\rm sr,h}\Phi_{{_{\rm h}}}\textbf{g}_{\rm rd,h}x_{\rm h} + n_{\rm h},\nonumber \\
		y_{\rm l} &= \sqrt{P_{\rm l}}\textbf{g}_{\rm sr,l}\Phi_{{\rm l}}\textbf{g}_{\rm rd,l}x_{\rm l} + n_{\rm l},
	\end{align}
	\noindent where $\mathbf{g}_{\rm sr,h} \in C^{1 \times N}$ and $\mathbf{g}_{\rm sr,l} \in C^{1 \times N}$ denote the channel vectors between the AP and the DBRIS at $f_{\rm h}$ and $f_{\rm l}$, respectively, $\mathbf{g}_{\rm rd,h} \in C^{N \times 1}$ and $\textbf{g}_{\rm rd,l} \in C^{N \times 1}$ represent the channel vectors between the DBRIS and the receivers operating at $f_{\rm h}$ and $f_{\rm l}$, respectively, the terms $n_{\rm h}$ and $n_{\rm l}$ denote the additive white Gaussian noise (AWGN) components affecting the system at the corresponding frequency bands, and where $\Phi_{{\rm h}}$ and $\Phi_{\rm l}$ are the response sub-matrices of the DBRIS defined earlier in section II. It is noteworthy that this system model can be extended to multi-user scenarios with one group of receivers in each frequency band.
	
	The channels are modeled to incorporate large-scale fading, characterized by path loss, and small-scale fading due to multipath propagation. The entries of the random vectors, i.e., $\mathbf{g}_{\rm sr,h}$, $\mathbf{g}_{\rm sr,l}$, $\mathbf{g}_{\rm rd,h}$ and $\mathbf{g}_{\rm rd,l}$, are modeled as independent and identically distributed (i.i.d.) random variables, following ${\mathcal{CN}}\left(0, \varepsilon_{\rm sr}\Omega\right)$, ${\mathcal{CN}}\left(0, \varrho_{\rm sr}\Omega\right)$, ${\mathcal{CN}}\left(0, \varepsilon_{\rm rd}\Omega\right)$, and ${\mathcal{CN}}\left(0, \varrho_{\rm rd}\Omega\right)$, respectively, with $\Omega$ representing the average power of the signal envelope,\footnote{Without loss of generality, the Nakagami parameter $m$ is set to unity in the simulations.} and with $\varepsilon_{\rm sr}$, $ \varrho_{\rm sr}$, $\varepsilon_{\rm rd}$ and $ \varrho_{\rm rd}$ denoting the large-scale fading coefficients computed at frequency $f \in\{f_{\rm h}, f_{\rm l}\}$ using \cite{dvxav} as per
	\begin{equation}
		\epsilon = G_{\rm t} + G_{\rm r} - 26\log_{10}(f) -36.7 \log_{10} (d),
		\label{sd}
	\end{equation}
	\noindent where $\epsilon \in\{\varepsilon_{\rm sr}, \varrho_{\rm sr}, \varepsilon_{\rm rd}, \varrho_{\rm rd}\}$,  $G_{\rm t}$ and $G_{\rm r}$ represent the gain of the transmit and receive antenna pair in dBi, $f\in\{f_{\rm h}, f_{\rm l}\}$ denotes the operation frequency in GHz, and $d$ denotes the distance between the AP and the considered receiver in meters.
	\subsection{EE Maximization}
	\subsubsection{Problem Formulation}
	Our focus is on the critical objective of maximizing the EE of the proposed DBRIS-assisted communication system. As next-generation wireless systems are expected to deliver high data rates with reduced energy consumption, EE is a fundamental design metric, particularly in the context of dual-band systems where multiple transmission chains may operate simultaneously. To this end, we formulate the EE maximization problem, which inherently captures the trade-off between the system throughput and its total power consumption.
	
	The EE of the DBRIS-aided communication system is defined as
	\begin{align}
		\eta = \frac{R_{\rm tot}}{P_{\sum}},
		\label{eq47}
	\end{align}
	\noindent where $P_{\footnotesize\sum}$ denotes the system's total power consumption, given by $P_{\footnotesize \sum} = P_{\rm h} + P_{\rm l} + P_{\rm DBRIS} + P_{\rm AP} + P_{\rm UE}$.\footnote{Hereafter, $P_{\rm \max}$ refers to the maximum transmit power available at the AP for transmissions over the two frequency bands, i.e., $P_{\rm h} + P_{\rm l}$.} Here, $P_{\rm DBRIS}=P_{\rm c}+ N\times P_{\rm sc}$ expresses the total power consumed by the DBRIS, including its control circuitry power $P_{\rm c}$ and the power consumed by the surface elements, where $P_{\rm sc}$ is the power consumption per super-cell element, $P_{\rm AP}$ and $P_{\rm UE}$ denote the power consumed by the control circuitry at the AP and the communication receiver, i.e., the UE, respectively.\footnote{The exact values taken for the simulations will be detailed in the numerical results section.} 
    Besides, $P_{\rm h}$ and $P_{\rm l}$ refer to the transmit powers of the communication links operating over the high-frequency and low-frequency bands, respectively, and the sum-rate, in the unit of Kbit/s,
    is defined as
	\begin{align}
		R_{\rm tot} = \log_{2}\left(1 + \gamma_{\rm h}\right) + \log_{2}\left(1 + \gamma_{\rm l}\right),
		\label{sfsv}
	\end{align}
	\noindent where $\gamma_{\rm h} = \frac{P_{\rm h}\left|\mathbf{g}_{\rm sr,h}\mathbf{\Phi}_{{{\rm h}}}\mathbf{g}_{\rm rd,h}\right|^2}{\sigma_{\rm h}^2}$ and $\gamma_{\rm l} = \frac{P_{\rm l}\left|\mathbf{g}_{\rm sr,l}\mathbf{\Phi}_{{{\rm l}}}\mathbf{g}_{\rm rd,l}\right|^2}{\sigma_{\rm l}^2}$ are the signal-to-noise power ratios (SNRs) at the two information receivers, with $\sigma_{\rm h}^{2}$ and $\sigma_{\rm l}^{2}$ denoting the noise variances.
	
	The EE optimization problem is then formulated as follows:
	\begin{subequations}
		\begin{align}
			&\qquad\qquad\quad\underset{P_{\rm h},P_{\rm l}, \mathbf{\Phi}_{\rm h}, \mathbf{\Phi}_{\rm l}}{\rm {maximize}}\ \eta \nonumber\\
			&{\text{subject~to:}}\nonumber\\
			&{\quad \gamma_{\rm h} \geq \gamma_{\rm h}^{\rm th}}\\
			&{\quad \gamma_{\rm l} \geq \gamma_{\rm l}^{\rm th}}\\
			&{\quad P_{\rm h} + P_{\rm l} \leq P_{\rm \max}} \\
			&\quad\vert\beta_{{\rm h},i} e^{- \jmath \phi_{{\rm h},i}}\rvert = \lvert\beta_{{\rm l},i}e^{- \jmath \phi_{{\rm l},i}}\rvert= 1,~\forall i \in \left\{1, \cdots, N\right\},
		\end{align}
		\label{eq491}
\end{subequations}

\noindent where $P_{\rm \max}$ is the power budget of the AP, and where $\gamma_{\rm h}^{\rm th}$ and $\gamma_{\rm l}^{\rm th}$ are the SNR requirements of the end users. The formulated optimization problem is to be solved at the AP, which is assumed to have the required channel state information (CSI) as well as the most relevant system parameters.
	
	The above-formulated optimization is an NP-hard non-convex problem, due to the non-convex and fractional nature of the objective function w.r.t. $P_{\rm h}$, $P_{\rm l}$, $\mathbf{\Phi}_{\rm h}$ and $\mathbf{\Phi}_{\rm l}$. Therefore, it is challenging to obtain an analytical solution for it using traditional methods. To tackle this, we make use of the Dinkelbach method \cite{8314727}, and reformulate the objective function into a difference of numerator and quasi-denominator form.\footnote{Though a feasible solution can be obtained by using the Dinkelbach or Quadratic transforms, the Dinkelbach method is superior in terms of convergence \cite{8314727}.} Hence, the problem (\ref{eq491}) is recast as
	\begin{subequations}
		\begin{align}
			&\quad\underset{P_{\rm h},P_{\rm l}, \mathbf{\Phi}_{\rm h}, \mathbf{\Phi}_{\rm l}}{\rm {maximize}}\ {R_{\rm tot} - \eta\left(P_{\rm h} + P_{\rm l} + P_{\rm DBRIS} + P_{\rm AP} + P_{\rm UE}\right)} \nonumber\\
			&{\text{subject~to:}}\nonumber\\
			&\quad {\gamma_{\rm h} \geq \gamma_{\rm h}^{\rm th}}\\
			&\quad {\gamma_{\rm l} \geq \gamma_{\rm l}^{\rm th}}\\
			&\quad{P_{\rm h} + P_{\rm l} \leq P_{\rm \max}} \\
			&\quad{|\beta_{{\rm h},i} e^{- \jmath \phi_{{\rm h},i}}|=|\beta_{{\rm l},i} e^{- \jmath \phi_{{\rm l},i}}| = 1, \forall i \in \left\{1, \cdots, N\right\}.}
		\end{align}
		\label{eq49}
	\end{subequations}

It is clear that the optimization remains non-convex. To solve this problem, we adopt the alternating optimization approach. Specifically, first the transmit powers will be fixed and the problem will be solved w.r.t. the phase-shift matrices $\mathbf{\Phi}_{\rm h}$ and $\mathbf{\Phi}_{\rm l}$. Then, for fixed phase-shift matrices, a power allocation strategy satisfying the constraints of the optimization will be developed.
	\subsubsection{Proposed Solution}
	\subsubsection*{\bf Phase Shifting Design}
	In this step, the optimization problem reduces to
	\begin{subequations}
		\begin{align}
			&\qquad\underset{\mathbf{\Phi}_{\rm h}, \mathbf{\Phi}_{\rm l}}{\rm {maximize}}\ {R_{\rm tot}} \nonumber\\
			&{\text{subject~to:}}\nonumber\\
			&\quad {\gamma_{\rm h} \geq \gamma_{\rm h}^{\rm th}}\\
			&\quad {\gamma_{\rm l} \geq \gamma_{\rm l}^{\rm th}}\\
			&\quad {|\beta_{{\rm h},i} e^{- \jmath \phi_{{\rm h},i}}|=|\beta_{{\rm l},i} e^{- \jmath \phi_{{\rm l},i}}| = 1, \forall i \in \left\{1, \cdots, N\right\}.}
		\end{align}
		\label{eq493}
	\end{subequations}

This problem can also be tackled using alternating optimization techniques. Specifically, first we solve it for fixed $\mathbf{\Phi}_{\rm h}$, and then for fixed $\mathbf{\Phi}_{\rm l}$. With further mathematical manipulations, the two sub-problems involved are formulated as follows:
	\begin{subequations}
		\begin{align}
			&\qquad\underset{\mathbf{\Phi}_{\rm h}}{\rm {maximize}}\ {\left|\mathbf{g}_{\rm sr,h}\mathbf{\Phi}_{{{\rm h}}}\mathbf{g}_{\rm rd,h}\right|^2} \nonumber\\
			&{\text{subject~to:}}\nonumber\\
			&\qquad {\gamma_{\rm h} \geq \gamma_{\rm h}^{\rm th}}\\
			&\qquad {|\beta_{{\rm h},i}  e^{- \jmath \phi_{{\rm h},i}}| = 1,  \forall i \in \left\{1, \cdots, N\right\}.}
		\end{align}
		\label{eq19}
	\end{subequations}
	\begin{subequations}
		\begin{align}
			&\quad\underset{\mathbf{\Phi}_{\rm l}}{\rm {maximize}}\ {\left|\mathbf{g}_{\rm sr,l}\mathbf{\Phi}_{{{\rm l}}}\mathbf{g}_{\rm rd,l}\right|^2} \nonumber\\
			&{\text{subject~to:}}\nonumber\\
			&\quad {\gamma_{\rm l} \geq \gamma_{\rm l}^{\rm th}}\\
			&\quad {|\beta_{{\rm l},i}  e^{- \jmath \phi_{{\rm l},i}}| = 1, \forall i \in \left\{1, \cdots, N\right\}.}
		\end{align}
		\label{eq20}
	\end{subequations}

The phase shifts follow a $\textit{S}^1$ or a one-dimensional manifold structure \cite{10783776}, according to
	\begin{align}
		\textit{S}^1_{f_{\rm l}} &= \left\{e^{-\jmath \phi_{{\rm l},i}} \in \textsc{C}~|~\phi_{{\rm l},i} \in [0, 2 \pi),~ i=1,\cdots,N\right\},\\
		\textit{S}^1_{f_{\rm h}} &= \left\{e^{-\jmath \phi_{{\rm h},i}} \in \textsc{C}~|~\phi_{{\rm h},i} \in [0, 2 \pi),~ i=1,\cdots,N\right\},
	\end{align}
	\noindent where each element response of the DBRIS belongs to $\textsc{C}$, which denotes the complex circle or the $S^1$ manifold structure.
	
	Initializing with a feasible set $\boldsymbol{\phi}_{\rm h}^{0} = [\phi_{{\rm h},1}^{0}, \cdots, \phi_{{\rm h},N}^{0}]$, the gradient descent w.r.t. $\mathbf{\phi}_{\rm h}$, is computed using
\begin{align}
		\mathbf{\phi}_{{\rm h}}^{k+1} = \mathbf{\phi}_{\rm h}^{k} + \alpha \nabla_{M} z_{\rm h}(\mathbf{\phi}_{{\rm h}}^{k}),
		\label{eq24}
	\end{align}
	where $z_{\rm h}(\mathbf{\phi}_{{\rm h}}^{k})=\left|\mathbf{g}_{\rm sr,h}\mathbf{\Phi}_{{{\rm h}}}^{\rm k}\mathbf{g}_{\rm rd,h}\right|^2 + \varphi$ is the modified objective function of problem (\ref{eq19}), $\mathbf{\Phi}_{\rm h}^{k}={\rm diag}\left(\mathbf{\phi}_{{\rm h},1}^{k},\cdots,\mathbf{\phi}_{{\rm h},N}^{k}\right)$, $\alpha$ denotes the step size,  $\varphi\left(k\right) = \max \left(0, \rho_{\rm pen} (\gamma_{\rm h}^{\rm th} - \gamma_{\rm h}^{k})\right)$ is the penalty term imposed when the constraints (\ref{eq19}a) and  (\ref{eq20}a) are violated at the $k^{\rm th}$ iteration, and $\nabla_{M}$ is the Riemannian gradient at iteration $k$.\footnote{For simplicity, we considered that the penalty factor $\rho_{\rm pen}$ is constant. This reduces the complexity for the implementation of the algorithm.} For the purpose of tracing the updated optimization variables $\mathbf{\phi}_{{\rm h}}^{k+1}$ back to the manifold structure, we make use of a Retraction operator defined as
\begin{align}
		{\rm \mathbf{Retr}}_{\textit{S}_{f_{\rm h}}^1} =  \frac{\mathbf{\phi}_{{\rm h}}^{k+1}}{\big|\mathbf{\phi}_{{\rm h}}^{k+1}\big|}.
		\label{eq25}
	\end{align}

The computation of the gradient and the back-projection operation can be performed
in an iterative way for $K_{\rm \max}$ times or until convergence. To enforce the constraints, a barrier method is incorporated into the modified objective function as defined earlier, penalizing any constraint violation.
	
A process similar to the one described above is performed to obtain the phase-shift matrix corresponding to the operation frequency $f_{\rm l}$, i.e., $\mathbf{\Phi}_{\rm l}^{\rm opt}$.
	
	The specific steps for obtaining the optimal phase-shift matrices with fixed transmit powers $P_{\rm h}$ and $P_{\rm l}$ are summarized in Algorithm 1.
	
	\begin{algorithm}[H]
		\caption{Phase-shift design using manifold optimization.}
		\begin{algorithmic}[1]
			\renewcommand{\algorithmicrequire}{\textbf{Input:}}
			\renewcommand{\algorithmicensure}{\textbf{Output:}}
			\REQUIRE $N$, $d_{\rm ru_{l}}$, $d_{\rm ru_{h}}$, $\gamma_{\rm l}^{\rm th}$, $\gamma_{\rm h}^{\rm th}$, $P_{\rm h}$, $P_{\rm l}$, $f_{\rm h}$, $f_{\rm l}$, $\vartheta_1$, $K_{\rm \max}$.
			\\
            \STATE Set $k = 0$.
			\STATE \textit{Generate the channel data, $\textbf{g}_{\rm sr,h}$, $\textbf{g}_{\rm sr,l}$, $\textbf{g}_{\rm rd,h}$, $\textbf{g}_{\rm rd,l}$}.\\
			\STATE  Initialize $\phi_{\rm h}^{0} = [\phi_{\rm h,1}^{0}, \cdots, \phi_{\rm h,\textit{N}}^{0} ]$. \\
			\STATE \textbf{Repeat} \\
            \hangindent=2.3em \hangafter=0
			\quad Perform gradient descent as per \eqref{eq24}. \\
			\quad Update the point on the manifold surface as per \eqref{eq25}. \\
			\quad Update $k = k+1$.
			\STATE \textbf{Until} $\left(\phi_{\rm h}^{k} - \phi_{\rm h}^{k-1}\right) \leq \vartheta_1$ or $k=K_{\max}$.
			\STATE Repeat steps (2)--(6) for $\mathbf{\Phi_{\rm l}}$.
			\ENSURE  $\mathbf{\Phi}_{\rm l}^{\rm opt} =\mathbf{\Phi}_{\rm h}^{k}$ and $\mathbf{\Phi}_{\rm h}^{\rm opt}=\mathbf{\Phi}_{\rm h}^{k}$.
		\end{algorithmic}
            \label{Algorithmdc2}
	\end{algorithm}
	\subsubsection*{\bf Power Allocation Strategy}
	Here, the optimization problem is formulated as follows:
	\begin{subequations}
		\begin{align}
			&\qquad\underset{P_{\rm h},P_{\rm l}}{\rm {maximize}}\  \eta \nonumber\\
			&{\text{subject~to:}}\nonumber\\
			&\qquad {\gamma_{\rm h} \geq \gamma_{\rm h}^{\rm th}}\\
			&\qquad {\gamma_{\rm l} \geq \gamma_{\rm l}^{\rm th}}\\
			&\qquad{P_{\rm h} + P_{\rm l} \leq P_{\rm \max}}.
		\end{align}
		\label{eq4923}
	\end{subequations}

Since the quality-of-service constraints in (\ref{eq4923}a) and (\ref{eq4923}b) are non-convex, it is difficult to solve the problem directly. To deal with this, the Taylor series expansion of the SNRs is leveraged to transform the said constraints into convex ones \cite[Eq. (12)]{10757325}. The resulting approximated SNRs are denoted by $\tilde{\gamma}_{\rm l}\geq \gamma_{\rm l}^{\rm th}$ and $\tilde{\gamma}_{\rm h}\geq \gamma_{\rm h}^{\rm th}$. The constraints are explicitly lower bounded using Taylor series approximations to a convex form in the subsequent sections as well.

Specifically, let $\mathbf{v} \triangleq \mathrm{diag}\{\mathbf{\phi}_i\}$, $\mathbf{P} \triangleq P_{i} \mathbf{g}_i \mathbf{g}_i^{\mathrm{H}}$, and $\mathbf{g}_i \triangleq \mathrm{diag}\{\mathbf{G}_{\mathrm{sr},i}^{\mathrm{H}}\} \, \mathbf{g}_{\mathrm{rd},i}$, where \( i \in \{{\mathrm{h}}, {\mathrm{l}}\} \) indicates the frequency band. The Taylor series expansion of the constraints (\ref{eq4923}a) and (\ref{eq4923}b) can be expressed as $\tilde{\gamma} \left(\mathbf{v}_t\right)= \mathbf{v}_{t-1}^H\textbf{P}\mathbf{v}_{t-1} + 2(\textbf{P}\mathbf{v}_{t-1})^H\left( \mathbf{v}_t - \mathbf{v}_{t-1} \right)$. Here, $t$ represents the iteration number. The problem still poses non-convexity due to the interlink between the variables $\eta$, $P_{\rm h}$, and $P_{\rm l}$. A feasible solution can be proposed using the alternating technique after the Dinkelbach transformation of the objective function \cite{8314727}.
	
	The optimal EE level, $\eta^{\rm opt}$, can be obtained using the Dinkelbach method through an effective iterative algorithm which updates $\eta$ at each iteration until it converges to $\eta^{\rm opt}$ or the maximum number of iterations is reached. The proposed procedure is summarized in Algorithm 2.
	
	\begin{algorithm}
		\caption{Iterative algorithm for maximizing EE.}
		\begin{algorithmic}[1]
			\renewcommand{\algorithmicrequire}{\textbf{Input:}}
			\renewcommand{\algorithmicensure}{\textbf{Output:}}
			\REQUIRE $P_{\rm \max}$, $f_{\rm h}$, $f_{\rm l}$, $\vartheta_2$, $\eta_{\rm init}$, $T_{\max}$.
			\\
			\STATE Set $t=0$.\\
          \STATE Solve problems (\ref{eq19}) and (\ref{eq20}) to obtain $\mathbf{\Phi}_{\rm l}^{\rm opt}$ and $\mathbf{\Phi}_{\rm h}^{\rm opt}$ using Algorithm~\ref{Algorithmdc2}.\\
			\STATE \textbf{Repeat}\\
			 \quad Given $\eta^t$, solve (\refeq{eq4924}) for $\mathbf{p}^t$ using Algorithm~\ref{Algorithmdc1}.\\
			\quad Set $\eta^{t} = \frac{\log(1 +\tilde{\gamma}_{\rm h}^{t}) + \log(1 +\tilde{\gamma}_{\rm l}^{t})}{P_{\rm h}^{t} + P_{\rm l}^{t} + P_{\rm DBRIS} + P_{\rm AP}+P_{\rm UE}}$ .\\
			\quad Update $t=t+1$.\\
			\STATE \textbf{Until} $\vert\eta^{t} -\eta^{t-1}\rvert \leq \vartheta_2$ or $t=T_{\max}$.\\
			\ENSURE $\eta^{\rm opt}=\eta^{t}$,  $P_{\rm h}^{\rm opt}=P_{\rm h}^{t}$ and $P_{\rm l}^{\rm opt}=P_{\rm l}^{t}$.
		\end{algorithmic}
	\end{algorithm}
	
	As shown in Algorithm 2, at each iteration $t$, we need to solve the following optimization problem for a given $\eta^{t}$:
	\begin{subequations}
		\begin{align}
			&\qquad\underset{P_{\rm h},P_{\rm l}}{\rm {maximize}}\  {R_{\rm tot}^{t} - \eta^{t}\left(P_{\rm h}^{t} + P_{\rm l}^{t} + P_{\rm DBRIS} + P_{\rm AP}+P_{\rm UE}\right)} \nonumber\\
			&{\text{subject~to:}}\nonumber\\
			&\qquad {\tilde{\gamma}_{\rm h} \geq \gamma_{\rm h}^{\rm th}}\\
			&\qquad {\tilde{\gamma}_{\rm l} \geq \gamma_{\rm l}^{\rm th}}\\
			&\qquad{P_{\rm h} + P_{\rm l} \leq P_{\rm \max}} \\
			&\qquad{P_{\rm h} \geq 0,~P_{\rm l} \geq 0}.
		\end{align}
		\label{eq4924}
	\end{subequations}
	To this end, the objective function is first recast as
	\begin{align}
		f(\mathbf{p})&=\underbrace{\log_2\left(1 + \frac{P_{\rm h} A_{\rm h}}{\sigma_{\rm h}^2} \right) + \log_2\left(1 + \frac{P_{\rm l} A_{\rm l}}{\sigma_{\rm l}^2} \right)}_{\text{concave}} \nonumber\\
		&- \underbrace{\eta^t\left(P_{\rm h} + P_{\rm l} + P_{\rm DBRIS} + P_{\rm AP} + P_{\rm UE}\right)}_{\text{convex}},
	\end{align}
	where $A_{\rm h} = \left| \mathbf{g}_{\rm sr,h} \mathbf{\Phi}_{\rm h} \mathbf{g}_{\rm rd,h} \right|^2 $ , $A_{\rm l} = \left| \mathbf{g}_{\rm sr,l} \mathbf{\Phi}_{\rm l} \mathbf{g}_{\rm rd,l} \right|^2$, and $\mathbf{p}=[P_{\rm h}, P_{\rm l}]$. The functional $f(\mathbf{p})$ is the difference between a concave function and a convex one, hence tractable by using DC programming for fixed values of $A_{\rm h}$ and $A_{\rm l}$, or for fixed values of the elements of $\mathbf{\Phi}_{\rm h}$ and $\mathbf{\Phi}_{\rm l}$. Therefore, the problem (\refeq{eq4924}) can be solved using DC programming iterative techniques.

The proposed power allocation method is presented in Algorithm 3.
	
	\begin{algorithm}
		\caption{Power allocation based on DC programming.}
		\begin{algorithmic}[1]
			\renewcommand{\algorithmicrequire}{\textbf{Input:}}
			\renewcommand{\algorithmicensure}{\textbf{Output:}}
			\REQUIRE$J_{\rm \max}$, $\vartheta_3$, $\mathbf{\Phi}_{\rm l}^{\rm opt}$, $\mathbf{\Phi}_{\rm h}^{\rm opt}$, $\mathbf{p}^{0}$.
			\\
            \STATE Set $j=0$.\\
			\STATE Calculate $f^{j}(\mathbf{p}^{j})$.\\
			\STATE \textbf{Repeat}\\
			\quad Solve (\refeq{eq4924}) using CVX for fixed  $\mathbf{\Phi}_{\rm l}^{\rm opt}$, $\mathbf{\Phi}_{\rm h}^{\rm opt}$ and $\eta^{t}$.\\
			\quad Set $\mathbf{p}^{j}=\mathbf{p}^{\rm opt}$ and calculate $f^{j}(\mathbf{p}^{j})$. \\
            \quad Update $j=j+1$.\\
			\STATE \textbf{Until} $\lvert f^{j}(\mathbf{p}^{j})-f^{j-1}(\mathbf{p}^{j})\rvert\leq \vartheta_3$ or $j=J_{\rm \max}$.\\
			\ENSURE  $\mathbf{p}^{j}=[P_{\rm h}^{j}, P_{\rm l}^{j}]$.
		\end{algorithmic}
        \label{Algorithmdc1}
	\end{algorithm}
	\subsection{Complexity Analysis}
	For completeness, we also analyze the computational complexity of the optimization methodology used to maximize EE in the DBRIS-assisted communication system.

The complexity of Algorithm 1 resides essentially in the manifold optimization operations and results in a computational cost of $\mathcal{O}(K_{\rm \max} N)$, which also accounts for the channel and matrix initialization steps. Recall that $K_{\rm \max}$ is the maximum number of iterations for the convergence of the algorithm. The complexity of Algorithm 2, on the other hand, stems from solving the convex subproblem in \eqref{eq4924}, which can be addressed in $\mathcal{O}(1)$ using standard convex optimization solvers such as CVX, typically based on interior-point methods. As this subproblem is solved at each iteration, the overall complexity of Algorithm 2 is $\mathcal{O}(T_{\max})$. Similarly, the complexity of Algorithm 3 is observed to be $\mathcal{O}(J_{\rm \max})$, where $J_{\rm \max}$ denotes the maximum number of iterations allowed \cite{cormen2009introduction}.
	\section{DBRIS Assisted SWIPT}
	\subsection{System Model}
The second use case under consideration is a SWIPT system comprised of a common AP and two spatially separated receivers as depicted in Fig. \ref{figa1}. In the considered SWIPT architecture, the lower frequency band ($f_{\rm l}$) is assigned to the wireless power transfer (WPT), while the higher frequency band ($f_{\rm h}$) is used for the information transmission. Allocating the lower frequency to WPT is generally preferred because it reduces the propagation loss and improves the RF-to-DC conversion efficiency. This higher conversion efficiency is largely due to reduced device parasitic effects and the use of simpler, less lossy impedance-matching networks in the receiver rectennas [41]. In contrast, operating WPT at higher frequencies amplifies parasitic losses, exacerbates rectifier nonlinearities, and complicates impedance matching, thereby increasing hardware complexity and reducing energy-harvesting performance [40], [41]. Consequently, assigning the higher frequency band to information transmission and the lower frequency band to WPT represents a practical and efficient design choice. The channel model described in section III is utilized for this application case as well. This SWIPT model can also be extended to multi-user scenarios in the future. As compared to existing studies on RIS-aided SWIPT, which are limited to a single band of operation, the proposed design leverages a DBRIS to enable simultaneous energy transfer and information transmission over distinct frequency bands, where the low frequency is leveraged to ensure improved wireless power transfer efficiency and, hence, higher levels of harvested energy at the target device.
	
The SWIPT system performance is evaluated using an EE utility function, which characterizes the trade-off between the achievable information rate, the harvested energy, and the transmit power \cite{10718349, s21186013}. This metric is widely adopted in the SWIPT literature. To this end, we formulate a joint utility function maximization problem that simultaneously considers the three objectives.
	
	Adopting the practical sigmoid-based non-linear energy harvesting model \cite{s21186013, 8892587}, the harvested energy and the achievable data rate are formulated as follows:
\begin{align}
		E &=\frac{E_{\rm max}}{1 - \zeta}
		\left(\frac{1}{1 + \exp \left(-\varsigma a\left(P_{\rm l} A_{\rm l} + b\right)\right)}- \zeta\right),
        \label{eq30a}
        \\
		R_{\rm h} &= \log_{2}\left(1 + (1 - \varsigma )\gamma_{\rm h}\right),
		\label{eq30b}
	\end{align}
	\noindent where $\zeta = \left(1 + \exp\left(ab\right)\right)^{-1}$, $a$ and $b$ are parameters related to the characteristics of the capacitor and the diode turn-on voltage in the energy harvesting circuitry, the values of which can be empirically determined using curve-fitting techniques as described in \cite{s21186013}, parameter $\varsigma$ denotes the power splitting factor between the two frequency bands, and $E_{\rm max}$ represents the maximum level of harvested energy when the harvesting circuit reaches saturation. Additionally, $\gamma_{\rm h}$ and $P_{\rm l} A_{\rm l}$ denote the received SNR at the communication device and the harvested power at the energy harvesting device, respectively.
	
	Accordingly, the weighted utility function, also referred to as the EE utility function \cite{10718349}, is defined as
	\begin{align}
		\eta_{\rm sw} = \frac{w_{\rm h} R_{\rm h} + w_{\rm l} E}{w_{\rm p} P_{\rm \sum}},
		\label{dfbno}
	\end{align}
	where $w_{\rm h}$, $w_{\rm l}$ and $w_{\rm p}$ are the weights of the objective function, and $P_{\rm \sum} = P_{\rm h} + P_{\rm l} + P_{\rm DBRIS} + P_{\rm AP} + P_{\rm UE}$ denotes the total power consumption of the system. The terms in $P_{\rm \sum}$ have been defined in section III, along with their descriptions.
	\subsection{Utility Function Maximization}
	\subsubsection{Problem Formulation}	
	Although utility functions such as the one in (\ref{dfbno}) have been widely considered in the literature as key metrics for evaluating EE, to the best of the authors' knowledge, this work is the first to apply such a utility function to assess the performance of the DBRIS-assisted SWIPT system design proposed in this paper.

Based on the utility function presented above, the optimization problem is formulated as follows:
	\begin{subequations}
		\begin{align}
			&\qquad\underset{P_{\rm h}, P_{\rm l}, \mathbf{\Phi}_{\rm h}, \mathbf{\Phi}_{\rm l}, \varsigma}{\rm {maximize}}\
			\eta_{\rm sw} \nonumber
			\\
			&{\text{subject~to:}}\nonumber\\
			& \quad {\gamma_{\rm h} \geq \gamma_{\rm h}^{\rm th}},  \\
			& \quad {E \geq E^{\rm th}}, \\
			& \quad {P_{\rm h}+P_{\rm l} \leq P_{\rm \max}},  \\
			& \quad 0 \leq \varsigma\leq 1, \\
			&\quad {|\beta_{{\rm h},i} e^{- \jmath \phi_{{\rm h},i}}|=|\beta_{{\rm l},i} e^{- \jmath \phi_{{\rm l},i}}| = 1, \forall i \in \left\{1, \cdots, N\right\}}, \\
			& \quad {w_{\rm h} + w_{\rm l}+ w_{\rm p}= 1},
		\end{align}
		\label{eqzdvz}
	\end{subequations}
	\noindent where $E^{\rm th}$ denotes the minimum level of energy according to the requirement of the harvesting device as defined earlier in \eqref{eq30a}.
	
	The non-convex nature of this problem presents challenges in obtaining a globally optimal solution through joint optimization of the decision variables. To address this, we assume that the weights, i.e., $w_{\rm h}$, $w_{\rm l}$, and  $w_{\rm p}$, are fixed according to predefined preferences. By systematically varying these weights, the Pareto-optimal front of the problem can be obtained, thereby capturing the trade-offs among the competing objectives.\footnote{The weights are constrained to sum up to unity, a condition that facilitates the convergence of the proposed algorithm. Investigation of the values of the weights for specific operational objectives is beyond the scope of this paper.}
	
	To address the challenges of the fractional programming problem at hand, we employ the Dinkelbach transformation \cite{cxbnxnx}, which has proven effective in handling similar cases. The approach consists of reformulating the original objective function from its fractional form into a subtractive form, representing the difference between the numerator and a scaled quasi-denominator, thereby facilitating tractable optimization.
	\subsubsection{Proposed Solution}	
The solution methodology follows an alternating optimization approach, in which the original problem is decomposed into two subproblems: one for optimizing the transmit power allocation, and the other for tuning the phase shifts of the DBRIS elements \cite{s21186013}. Each of these subproblems falls within the framework of disciplined convex-concave programming (DCCP).

Additionally, the harvested energy can be further transformed into an equivalent data rate \cite{8854318, Zina2025}. This transformation is motivated by the practical consideration that the receivers' 
harvested power is typically stored in a battery for subsequent use to support data transmission on the uplink. Accordingly, the harvested energy is re-expressed as
	\begin{align}
		E^{\prime} = \log_2\left(1 + \eta_{\rm con}E \right),
	\end{align}
where $\eta_{\rm con}$ denotes the energy conversion efficiency \cite{Xia2015} to ensure unit consistency between the data rate (in kbits/s) and the harvested power (in watts). The value of $\eta_{\rm con}$
lies between 0 and 1, depending on the RF-to-DC conversion circuitry and the hardware implementation of the energy harvesting module. As such, the modified objective function incorporates $E^{\prime}$ in the utility function.\footnote{The transformation was carried out to ensure compliance with the DCCP framework required by CVX.}
	
	The Dinkelbach transform-based alternating optimization algorithm is proposed as a solution for the problem with the modified objective function given by
	\begin{align}
		\eta_{\rm sw} = \frac{w_{\rm h}R_{\rm h} + w_{\rm l} E^{\prime}}{ w_{\rm p} P_{\rm \sum}}.
	\end{align}
	The optimization problem is hence formulated as
	\begin{subequations}
		\begin{align}
			&\qquad\underset{P_{\rm h}, P_{\rm l}, \mathbf{\Phi}_{\rm h}, \mathbf{\Phi}_{\rm l}, \varsigma}{\rm {maximize}}
			w_{\rm h} R_{\rm h} + w_{\rm e} E^{\prime} - \eta_{\rm sw} w_{\rm p} P_{\rm \sum}\nonumber\\
			&{\text{subject~to:}}\nonumber \\
			& \quad {\gamma_{\rm h} \geq \gamma_{\rm h}^{\rm th}}, \\
			& \quad {E \geq E^{\rm th}},\\
            & \quad {P_{\rm h}+P_{\rm l} \leq P_{\rm \max}},  \\
			& \quad {0 \leq \varsigma \leq 1}, \\
			& \quad {|\beta_{{\rm h},i} e^{- \jmath \phi_{{\rm h},i}}|=|\beta_{{\rm l},i} e^{- \jmath \phi_{{\rm l},i}}| = 1,~\forall i \in \left\{1, \cdots, N\right\}}.
			\label{eqzbv}
		\end{align}
	\end{subequations}

To solve the above problem, we employ an alternating optimization approach, as presented in Algorithm 4. Further details are provided in the next subsection.
	\subsubsection*{\bf Phase Shifting Design}
	First, the problem is solved to obtain the optimized phase shifts for fixed values of $P_{\rm l}$, $P_{\rm h}$ and $\zeta$, using manifold optimization. The specific sub-problems are as follows:
	\begin{subequations}
		\begin{align}
			&\qquad\underset{\mathbf{\Phi}_{\rm h}}{\rm {maximize}}\ {\left|\mathbf{g}_{\rm sr,h}\mathbf{\Phi}_{{{\rm h}}}\mathbf{g}_{\rm rd,h}\right|^2} \nonumber\\
			&{\text{subject~to:}}\nonumber\\
			&\qquad {\gamma_{\rm h} \geq \gamma_{\rm h}^{\rm th}}\\
			&\qquad{|\beta_{{\rm h},i} e^{- \jmath \phi_{{\rm h},i}}| = 1,~\forall i \in \left\{1, \cdots, N\right\},}
		\end{align}
		\label{eq37}
	\end{subequations}
	\begin{subequations}
		\begin{align}
			&\quad\underset{\mathbf{\Phi}_{\rm l}}{\rm {maximize}}\ {\left|\mathbf{g}_{\rm sr,l}\mathbf{\Phi}_{{{\rm l}}}\mathbf{g}_{\rm rd,l}\right|^2} \nonumber\\
			&{\text{subject~to:}}\nonumber\\
			&\quad {E \geq E^{\rm th}}\\
			&\quad{|\beta_{{\rm l},i} e^{- \jmath \phi_{{\rm l},i}}| = 1,~\forall i \in \left\{1, \cdots, N\right\}.}
		\end{align}
		\label{eq38}
	\end{subequations}

The problems formulated in (\refeq{eq37}) and (\refeq{eq38}) are solved using the manifold optimization technique. First, we initialize the values for ${\Phi}_{\rm h}$ and ${\Phi}_{\rm l}$, followed by computation of the Riemannian gradient on the modified objective function, i.e., the sum of the objective function and the penalty, and projecting it back to the manifold surface using retraction operation till convergence. A more detailed description of the same has already been provided in section III.
	\subsubsection*{\bf Power Splitting Factor Design}
	Given the values of $\eta_{\rm sw}$, $P_{\rm h}$ and $P_{\rm l}$, and the optimized phase-shift matrices $\mathbf{\Phi}_{\rm h}^{\rm opt}$ and $\mathbf{\Phi}_{\rm l}^{\rm opt}$, the goal now is to determine the value of $\varsigma$. This optimization problem, which is formulated as
	\begin{subequations}
		\begin{align}
			& \qquad \underset{\zeta}{\rm {maximize}} \ {w_{\rm h} R_{\rm h} + w_{\rm l} E^{\prime} - \eta_{\rm sw} w_{\rm p} P_{\rm \sum}}\nonumber\\
			& {\text{subject~to:}}\nonumber \\
			& \qquad{\gamma_{\rm h} \geq \gamma_{\rm h}^{\rm th}}, \\
			& \qquad{E \geq E^{\rm th}},\\
			& \qquad{0 \leq \zeta \leq 1},
			\label{eqzbv12a}
		\end{align}
	\end{subequations}
	can be solved using convex tools such as CVX \cite{cvxdcx}.
	\subsubsection*{\bf Power Allocation Strategy}
	Given the value of $\eta_{\rm sw}$, the optimized power splitting factor $\varsigma$, and the optimized phase-shift matrices $\mathbf{\Phi}_{\rm h}$ and $\mathbf{\Phi}_{\rm l}$, we tackle the problem for obtaining the values of $P_{\rm h}$ and $P_{\rm l}$. This optimization problem is formulated as
	\begin{subequations}
		\begin{align}
			&\quad \underset{P_{\rm l}, P_{\rm h}}{\rm {maximize}} \quad
			w_{\rm h} R_{\rm h} + w_{\rm l} E^{\prime} - \eta_{\rm sw} w_{\rm p} P_{\rm \sum}\nonumber\\
			&{\text{subject~to:}}\nonumber \\
			&\qquad{\gamma_{\rm h} \geq \gamma_{\rm h}^{\rm th}}, \\
			&\qquad{E \geq E^{\rm th}},\\
			&\qquad{P_{\rm h}+P_{\rm l}\leq P_{\rm \max}},
			\label{eqzbv12}
		\end{align}
	\end{subequations}
and can be effectively solved using the iterative DC programming approach in conjunction with CVX.
	
	\begin{algorithm}
		\caption{Utility function maximization.}
		\begin{algorithmic}[1]
			\renewcommand{\algorithmicrequire}{\textbf{Input:}}
			\renewcommand{\algorithmicensure}{\textbf{Output:}}
			\REQUIRE $N$, $d_{\rm ru_h}$, $d_{\rm ru_l}$, $\gamma_{\rm h}^{\rm th}$, $E^{\rm th}$, $P_{\rm \max}$, $f_{\rm h}$, $f_{\rm l}$, $\vartheta_4$, $T_{\max}$.\\
			\STATE \textit{Generate the channel data, $\textbf{g}_{\rm sr,h}$, $\textbf{g}_{\rm sr,l}$, $\textbf{g}_{\rm rd,h}$, $\textbf{g}_{\rm rd,l}$}.\\
            \STATE Set $t=0$.\\
            \STATE Initialize $\mathbf{\Phi}_{\rm h}^{t}$, $\mathbf{\Phi}_{\rm l}^{t}$, $P_{\rm h}^{t}$, $P_{\rm l}^{t}$, $\varsigma^{t}$. \\
			\STATE \textbf{Repeat} \\
			\quad Given  $P_{\rm h}^{t}$, $P_{\rm l}^{t}$, $\varsigma^{t}$, Solve (\refeq{eq37}) and (\refeq{eq38}) to obtain
            $\mathbf{\Phi}_{\rm h}^{t+1}$ \hspace{1in} and $\mathbf{\Phi}_{\rm l}^{t+1}$ using Manifold optimization.\\
			\quad For $\mathbf{\Phi}_{\rm h}^{t+1}$, $\mathbf{\Phi}_{\rm l}^{t+1}$, $P_{\rm h}^{t}$, $P_{\rm l}^{t}$, $\eta_{\rm sw}^{t}$, obtain $\varsigma^{t+1}$ using CVX.\\
			\quad For $\mathbf{\Phi}_{\rm h}^{t+1}$, $\mathbf{\Phi}_{\rm l}^{t+1}$, $\eta_{\rm sw}^{t}$, $\varsigma^{t+1}$, obtain  $P_{\rm h}^{t+1}$ \& $P_{\rm l}^{t+1}$ using\\
            \hspace{0.9em} CVX.\\
			\quad Set $\eta_{\rm sw}^{t+1} = \frac{w_{\rm h} R_{\rm h} + w_{\rm l} E}{w_{\rm p} \left(P_{\rm h}^{t+1} + P_{\rm l}^{t+1} + P_{\rm AP} + P_{\rm DBRIS} + P_{\rm UE}\right)}$.\\
            \quad Update $t = t+1$.
            \\
			\STATE \textbf{Until} $\lvert\eta_{\rm sw}^{t} - \eta_{\rm sw}^{t-1}\rvert \leq \vartheta_4$ or $t=T_{\max}$.\\
			\ENSURE  $\eta_{\rm sw}^{\rm opt} = \eta_{\rm sw}^{t}$, $P_{\rm h}^{\rm opt}=P_{\rm h}^{t}$, $P_{\rm l}^{\rm opt}=P_{\rm l}^{t}$, $\mathbf{\Phi}_{\rm h}^{\rm opt}=\mathbf{\Phi}_{\rm h}^{t}$ and $\mathbf{\Phi}_{\rm l}^{\rm opt}=\mathbf{\Phi}_{\rm l}^{t}$.
		\end{algorithmic}
	\end{algorithm}
	\subsection{Complexity Analysis}
	Once again, we analyze the computational complexity of the proposed solution method. The dominant computational complexity of Algorithm 4 arises from the manifold optimization of the DBRIS phase shifts under the unit-modulus constraints. This step incurs a computational cost of $\mathcal{O}(K_{\max} N)$, where $K_{\max}$ denotes the maximum number of iterations it takes for the algorithm to converge using manifold optimization (cf. Algorithm 1). The subsequent updates of the power splitting factor and the power allocation involve solving low-dimensional convex problems, each with a complexity of $\mathcal{O}(1)$. Hence, considering $T_{\max}$ outer iterations, the overall computational complexity of Algorithm~4 is $\mathcal{O}(T_{\max} K_{\max} N)$.
	\section{Performance Evaluation}
	In this section, we analyze the performance of the DBRIS-aided systems under study, i.e., communication and SWIPT, and investigate the impact of key system parameters on performance. The operation frequencies are set as $f_{\rm h}=3.6$ GHz and $f_{\rm l}=2.5$ GHz. Unless stated otherwise, the distance between the source AP and the DBRIS is considered as 10 m, the distance from the DBRIS to device $u_{\rm h}$ served at the higher frequency is 15 m, and the one to device $u_{\rm l}$ is set to 30 m.\footnote{Considering the range of DBRIS elements arranged in a uniform planar array configuration and utilized in the paper with the reference design of \cite{jhgnh}, the far-field distance would be around 5 m for the low operating frequency.} The circuit power of each device is $P_{\rm UE} = 10$ dBm, while for the AP, we considered $P_{\rm AP} = 20$ dBm \cite{9497709}. For the varactor-based drive circuit, the drive power for each DBRIS element is set to 0.43 W, while a Xylinx-based controller is used for the simulation with $P_{\rm c} = 4.8$ W \cite{jhgnh, 10480438}. For the SWIPT use case, the values of the constant parameters are as follows: $a = 2000,~b = 0.0012,~E^{\rm th} = 0.5$ mJ, and $E_{\rm max} = 6$ mJ. Also, $\vartheta_1=\vartheta_2=\vartheta_3=\vartheta_4=10^{-6}$ and $\eta_{\rm con} = 0.8$. For both use cases, the antenna gains of the AP, DBRIS, and receivers are considered as 25 dBi. The bandwidth of operation is considered to be 100 KHz for both use cases.

For comparisons, we consider the operation of either of the systems under consideration with two conventional SBRISs, each operating at one of the said frequencies. In the cases with SBRIS deployment, two separate controllers are required, each of which requires a fixed static power $P_{\rm c}$. The power consumed by the drive circuit remains the same as in the DBRIS case. For a fair comparison between the DBRIS-aided system and the SBRIS-aided baseline, an ideal metasurface is considered in the current work. Consequently, the surfaces are assumed to possess continuous phase-shifting capabilities, thereby omitting the hardware limitations of discrete phase resolution. Additionally, all the power-unit definitions are considered in the logarithmic scale.

Next, numerical results, generated while varying the main operation parameters to assess the system performance of the communication and SWIPT models under study, are analyzed. The results have been generated using 500 channel realizations in the simulations.
	\subsection{DBRIS-Assisted Communication}
\begin{figure}[ht!]
		\centering
		\vspace*{-0.1in}
        	\hspace*{-0.1in}
		\includegraphics[height=3in,width=3.75in]{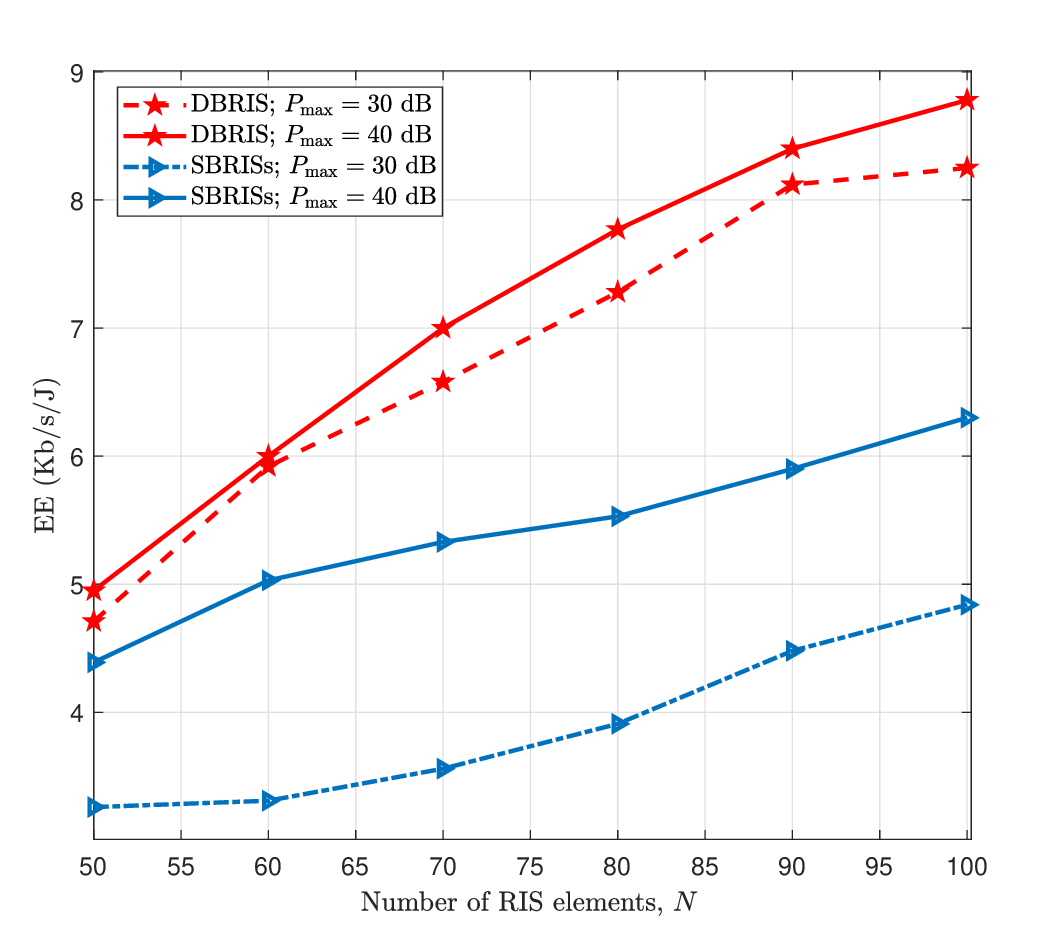}
		\vspace{-0.25in}
		\caption{EE versus the number of RIS elements, $N$.}
		\label{figqw1}
	\end{figure}

\begin{figure}[ht!]
		\centering
		\vspace{-0.2in}
		\hspace*{-0.1in}
\includegraphics[height=3in,width=3.85in]{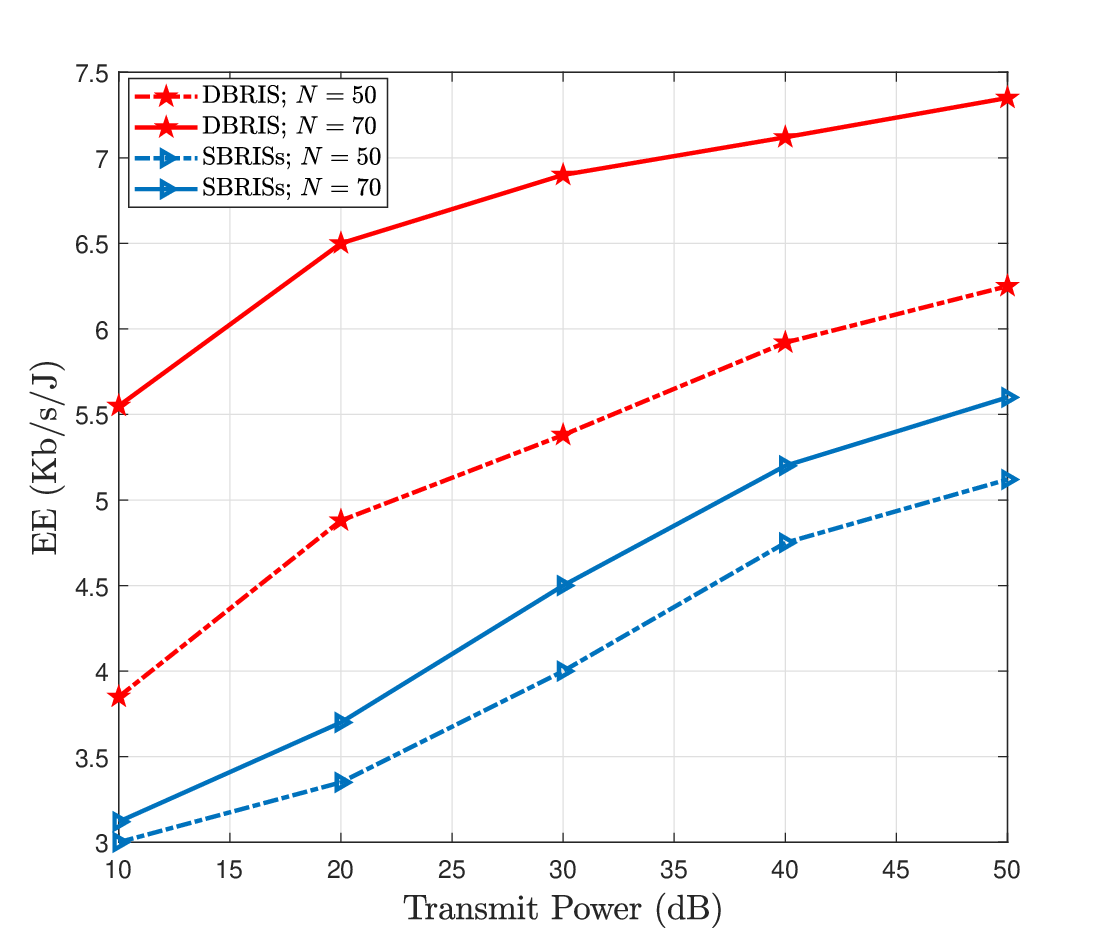}
		\vspace{-0.25in}
		\caption{EE versus the transmit power, $P_{\rm \max}$.}
		\label{figqw23}
	\end{figure}

	Figures \ref{figqw1} and \ref{figqw23} illustrate the EE gains achieved by the DBRIS-assisted communication system compared to its counterpart employing conventional SBRISs at the same location. As shown in Fig. \ref{figqw1}, increasing the maximum transmit power $P_{\max}$ leads to a consistent improvement in EE for both systems. However, DBRIS demonstrates a clear performance advantage, consistently outperforming the dual-SBRIS configuration across the considered power budgets (30 dB and 40 dB). This confirms the effectiveness of hardware sharing enabled by the DBRIS. For any fixed $P_{\rm \max}$, the EE scales almost linearly with the number of reflecting elements $N$, underscoring the importance of large surfaces in enhancing the passive beamforming gain. Complementary insights are provided in Fig.~\ref{figqw23}, where, for fixed $N$, increasing the AP's transmit power improves the EE. Among all configurations, DBRIS with $N=70$ achieves the highest EE across the entire transmit power range, thereby further demonstrating the superiority of the proposed architecture over SBRIS counterparts.

\begin{figure}[ht!]
		\centering
		\hspace*{-0.1in}
		 \includegraphics[height=3in,width=3.85in]{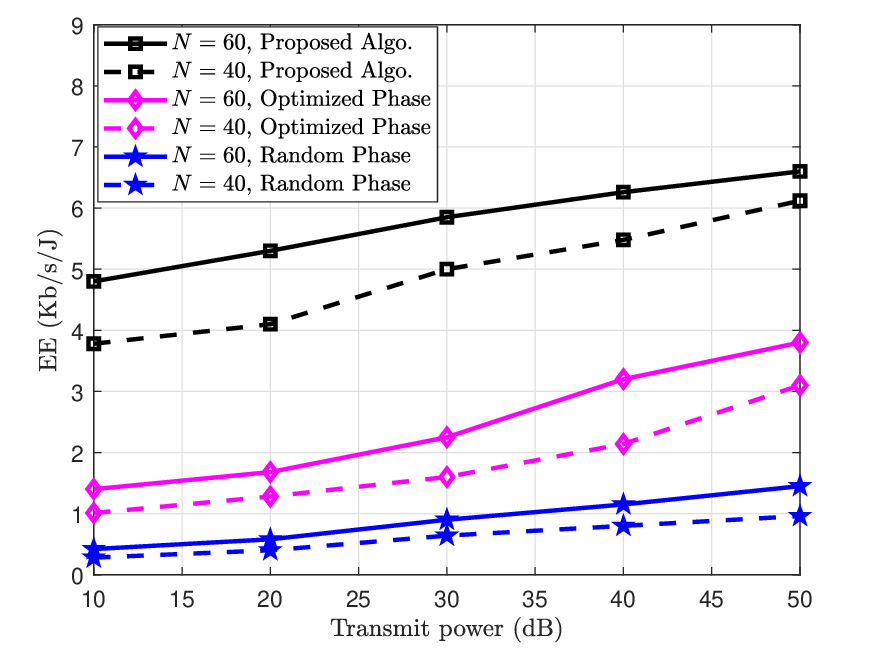}
		\vspace{-0.15in}
		\caption{Comparisons of the proposed algorithm against two distinct  baselines.}
		\label{fig11ab}
	\end{figure}

Figure~\ref{fig11ab} depicts
the superiority of the proposed algorithm over two commonly adopted baselines. The first one, referred to as `Optimized Phase' in the figure, employs an optimized phase configuration at the DBRIS with uniform power allocation at the AP across both operating bands. The second one, referred to as `Random Phase', considers random phase shifting at the DBRIS combined with optimal power allocation across the two operating bands at the AP. As depicted, the proposed algorithm outperforms both baselines, with the random phase setting performing the worst. This can be attributed to the fact that the phase alignment dictates the beamforming and the array gain, while power allocation is responsible for the amplitude scaling.
	
    	\begin{figure}[ht!]
		\centering
		\vspace{-0.2in}
		\hspace*{-0.1in}
\includegraphics[height=3in,width=3.85in]{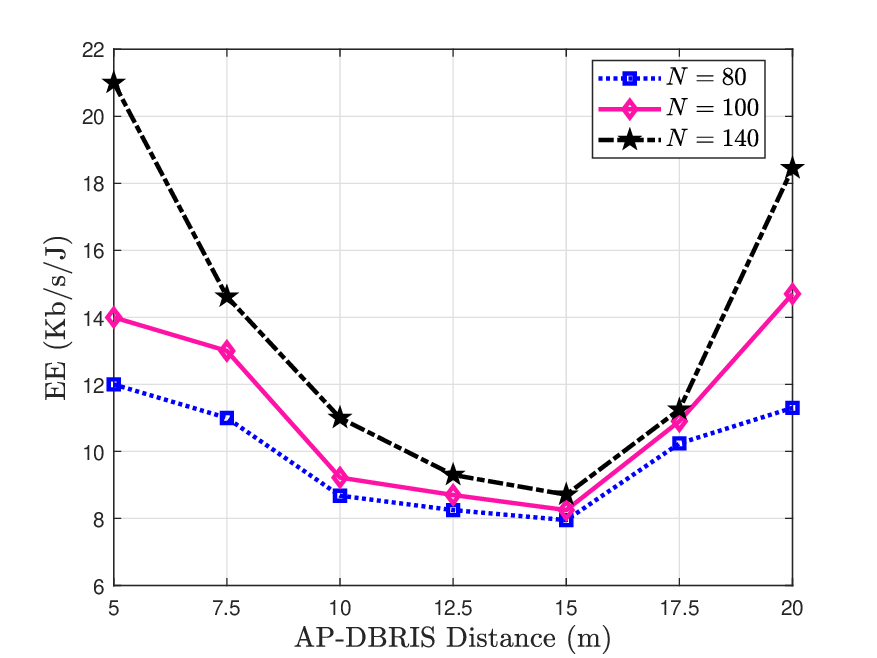}
		\vspace{-0.2in}
		\caption{EE versus the distance between the AP and the DBRIS, $P_{\max}= 40$ dB, \( d_{\rm h} = 25~\text{m} \) and \( d_{\rm l} = 35~\text{m}\).}
		\label{fig12d}
	\end{figure}

     Figure~\ref{fig12d} illustrates the variation of EE w.r.t. the distance between the AP and the center of the DBRIS for different values of $N$, while the separation between the DBRIS and the obstacle is fixed at $T=5$ m (cf. Fig.~\ref{figa1}). As expected, EE attains its maximum when the DBRIS is either close to AP or near the receivers, owing to stronger constructive reflections and reduced path loss in such cases. In contrast, placing the DBRIS midway between the AP and the users leads to reduced EE due to weaker received signals. Furthermore, increasing $N$ accentuates the variation in EE across different placements, as larger arrays provide stronger beamforming gains.

	\begin{figure}[ht!]
		\centering
        \hspace*{-0.15in}
		\includegraphics[height=3in,width=3.85in]{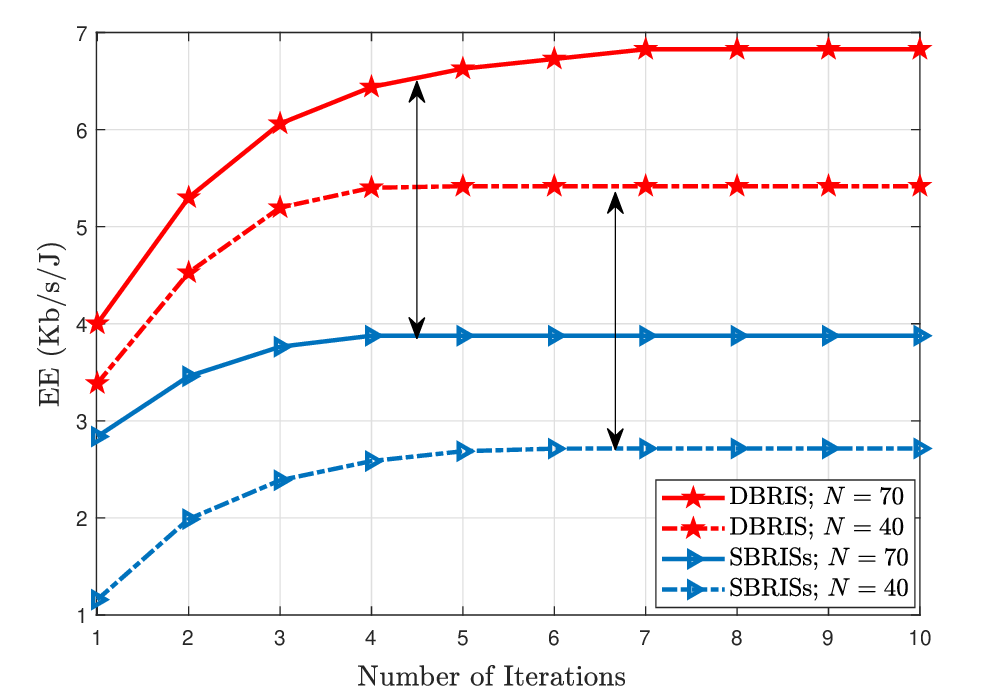}
		\vspace{-0.15in}
		\caption{EE versus the number of iterations for different values of $N$,$P_{\rm \max}=30$ dB.}
		\label{fig11}
	\end{figure}

Figure~\ref{fig11} shows the convergence behavior of Algorithm~2 for the DBRIS-aided system compared to a baseline with two SBRISs. The proposed algorithm converges to the optimal phase shifts and power allocations within seven iterations for both system models.

The results discussed above highlight DBRIS as a superior alternative to multiple SBRISs, delivering enhanced EE along with reduced hardware complexity and deployment cost. The adoption of multi-band metasurfaces not only facilitates their integration in different applications, as discussed earlier, but also minimizes the material usage and provides significant cost savings in manufacturing, installation, and operation. Collectively, these benefits underscore the merits of DBRIS from a sustainability standpoint.
	\subsection{DBRIS-Assisted SWIPT}
	
	\begin{figure}[h!]
		\centering
		\vspace{-0.1in}
        \hspace*{-0.1in}
		 \includegraphics[height=3in,width=3.8in]{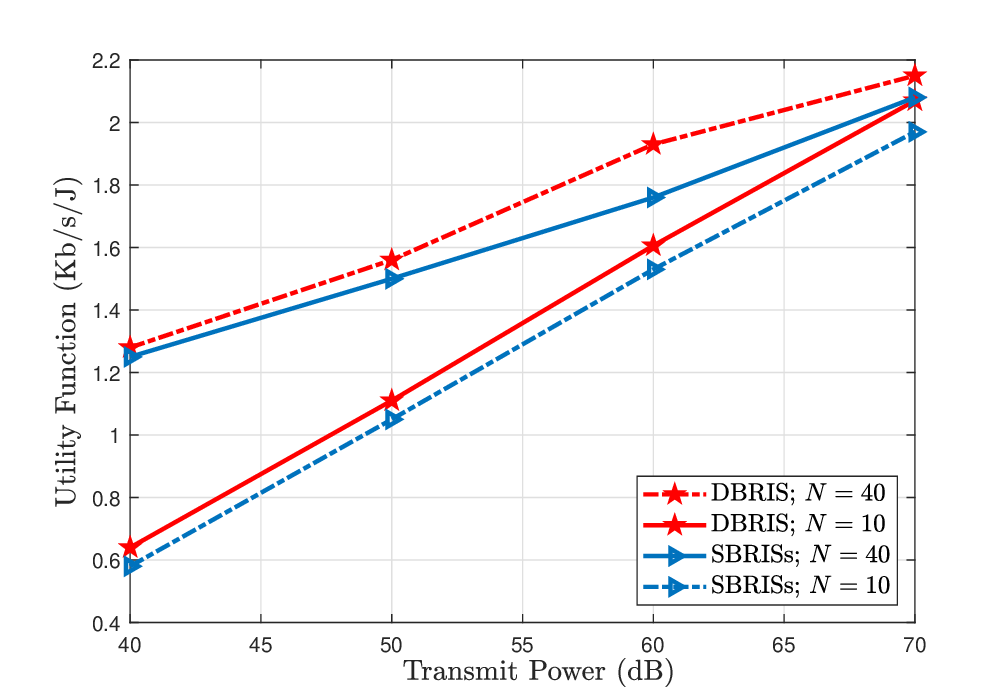}
		\vspace{-0.25in}
		\caption{Utility function versus the transmit power $P_{\max}$ for different DBRIS sizes. Here, $w_{\rm h}=w_{\rm l}=0.2$ and $w_{\rm p}=0.6$.}	
		\label{fig15sg}
	\end{figure}	
            	\begin{figure}[ht!]
		\centering
         \hspace*{-0.15in}
		 \includegraphics[height=3in,width=3.85in]{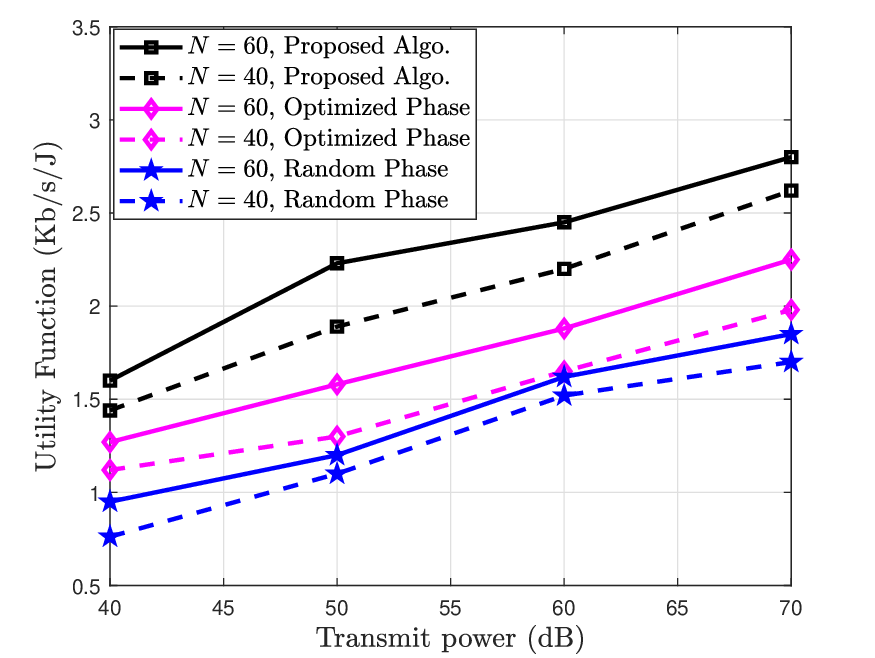}
		\vspace{-0.15in}
		\caption{Comparisons of the proposed algorithm against two distinct baselines for DBRIS-assisted SWIPT system.}
		\label{fig9ik}
	\end{figure}

	Figure~\ref{fig15sg} illustrates the variation of the utility function with the transmit power for different DBRIS sizes. The results show that increasing $N$ enhances the utility function for both the DBRIS-aided system and the SBRIS-assisted system. 
    Importantly, deploying a DBRIS yields additional gains compared to two separate SBRISs, owing to its simplified power requirement and reduced hardware overhead, including the elimination of extra circuitry, as discussed earlier.

Figure~\ref{fig9ik} illustrates the superiority of the proposed AO algorithm over two distinct baselines similar to the communication scenario. Recall that `Optimized Phase' and `Random Phase' baselines are defined earlier in the description of Fig.~\ref{fig11ab}.
Similar to the communication scenario, we observe that our proposed algorithm outperforms both baselines, with the random phase baseline being the worst. This can be attributed to the fact that the effect of the phase is more predominant compared to the amplitude for the DBRIS.

	\begin{figure}[h!]
		\centering
		\vspace{-0.02in}
         \hspace*{-0.1in}
		 \includegraphics[height=3in,width=3.85in]{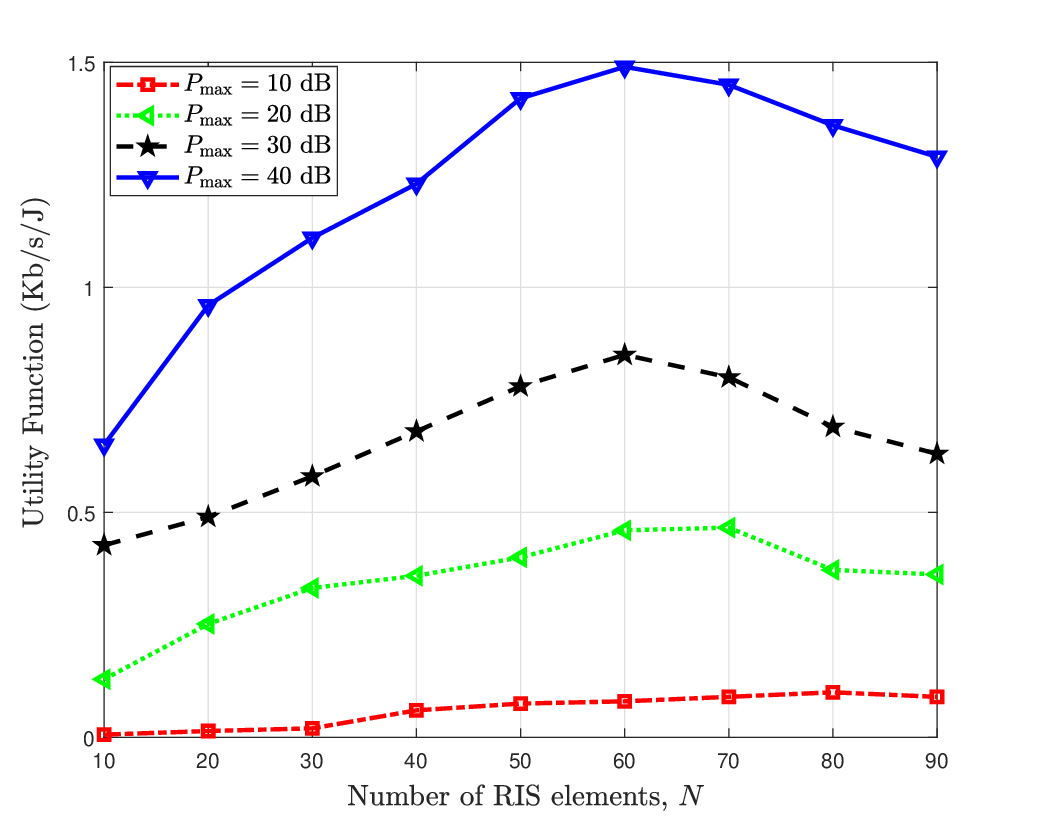}
		\vspace{-0.05in}
		\caption{Utility function versus the number of DBRIS elements for different values of the transmit power $P_{\max}$. Here, $w_{\rm h}=w_{\rm l}= 0.2$ and $w_{\rm p}=0.6$.}	
		\label{fig15bb}
	\end{figure}

Figure~\ref{fig15bb} shows the variation of the utility function w.r.t. the number of DBRIS elements for different values of the power budget $P_{\rm \max}$. As expected, the utility function improves as $P_{\rm \max}$ increases, reflecting the positive impact of higher available power on the system's ability to enhance the information and power transfer. For each power level, the utility function initially increases with $N$, confirming the gain provided by adding more DBRIS elements. A saturation or even a decline in utility function is noted beyond a certain DBRIS size, particularly beyond $N= 50$ for higher power budgets such as $P_{\rm \max} = 30~\text{dB}$ and $40~\text{dB}$. This signifies the presence of an inflection point, beyond which the additional DBRIS elements no longer yield gains in the system performance. This is attributed to the diminishing returns of the achievable sum-rate, which grows logarithmically with the signal power, in contrast to the linear growth in the power consumption associated with each added DBRIS element. The linear increase arises from the cumulative energy cost of biasing and controlling each DBRIS element, which becomes dominant as the surface size scales. While the raw RF power incident at the rectenna also scales with $N$, the SWIPT optimization must balance the allocation of transmit power across the communication and the WPT bands at the AP. The optimization framework dynamically routes the power allocation between the bands, causing the WPT curve to scale differently compared to the purely information-driven rate curve of Fig. \ref{figqw1}. In summary, the overall utility function value begins to decline beyond the optimal size, emphasizing the need for a careful joint optimization of the DBRIS size and the power allocation in the SWIPT scenario.

        \begin{figure}[h!]
		\centering
         \hspace*{-0.1in}
		\vspace{-0.1in}
        \includegraphics[height=3in,width=3.8in]{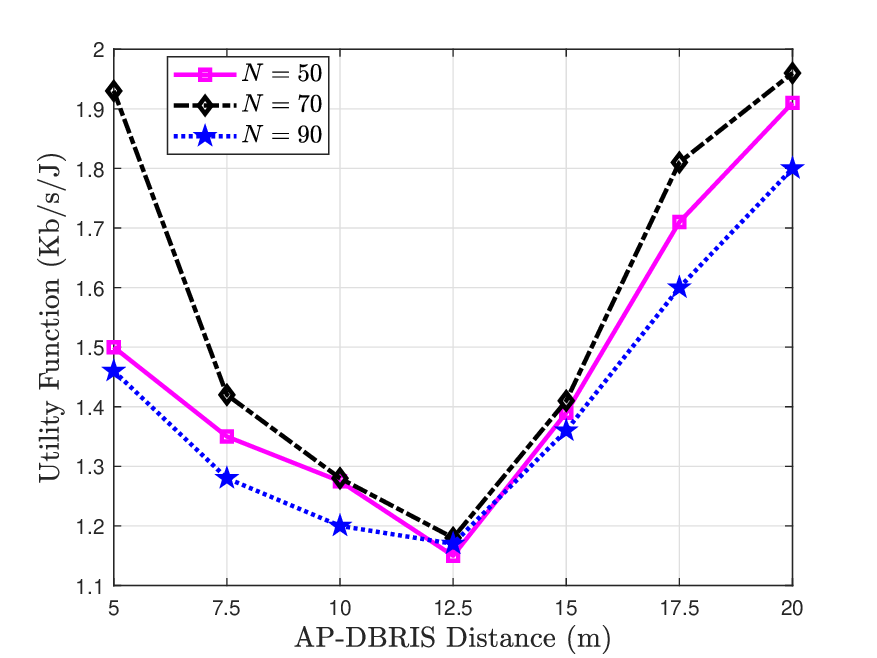}
		\vspace*{-0.2in}
		\caption{Utility function versus the distance between the AP and the DBRIS, $P_{\max}= 40$ dB, \( d_{\rm h} = 20~\text{m} \) and \( d_{\rm l} = 30~\text{m}\).}	
		\label{fig15}
	\end{figure}

    Figure~\ref{fig15} depicts the variation of the utility function as a function of the distance between the AP and the center of the DBRIS, for various values of $N$. The distance between the DBRIS and the obstacle, i.e., $T$, is held constant at 5~m (cf. Fig.~\ref{figa1}). As expected, the utility function reaches its peak when the DBRIS is either in close proximity to the transmitter or near the end receivers. This is attributed to the enhanced received signal power and improved harvested energy in these configurations, resulting from stronger constructive reflections and reduced path loss. Conversely, when the DBRIS is located midway between the AP and the users, the utility function diminishes due to weaker received signals caused by higher combined path loss. As observed in Fig. \ref{fig15bb} and from the law of diminishing returns, the utility function diminishes after reaching a maximum for an increase in the number of RIS elements. The variation of the utility function is also less compared to the variation of the EE in the DBRIS-assisted communication scenario for different values of RIS elements and AP-DBRIS distance.

	\begin{figure}[h!]
		\centering
         \hspace*{-0.1in}
		\includegraphics[height=3in,width=3.85in]{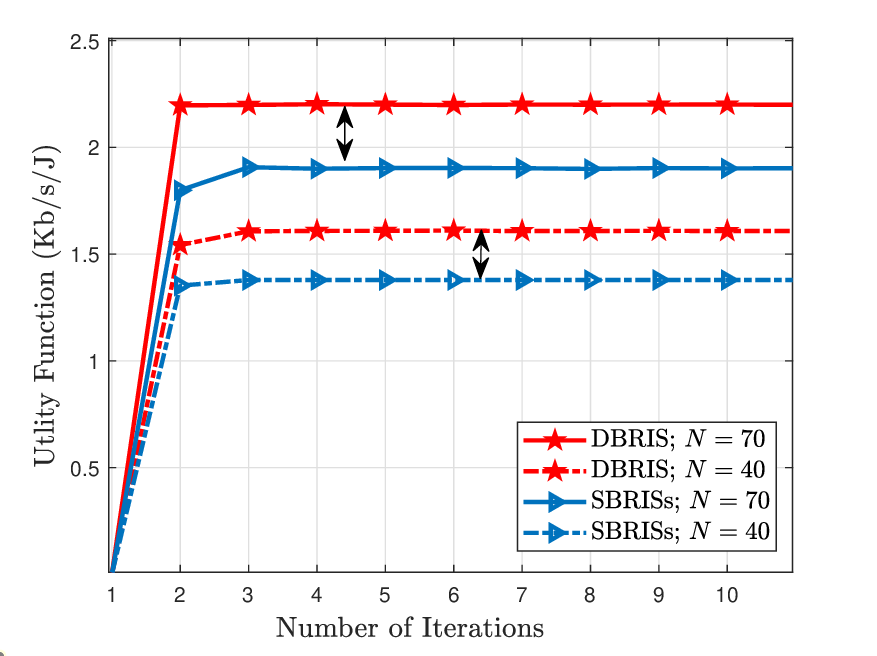}
		\caption{Utility function versus the number of iterations. Here, $\gamma_{\rm h}^{\rm th} = -50$ dB, $P_{\rm \max}=60$ dB, $w_{\rm h}=0.1,~w_{\rm l}=0.2$, and $w_{\rm p}=0.7$.}	
		\label{fig12}
	\end{figure}

Figure~\ref{fig12} presents the evolution of the utility function of the DBRIS-assisted SWIPT system versus the number of iterations, considering two values of $N$: $40$ and $70$. The performance of the system operation with two SBRISs of the same size is also shown. The results clearly indicate that the DBRIS-aided system achieves a higher utility function value than the SBRIS counterpart across all cases, owing to its ability to simultaneously serve users at different frequencies. Additionally, increasing the number of RIS elements enhances the performance, with a more significant gain observed in the DBRIS configuration. This is attributed to the increased beamforming resolution and improved energy focusing enabled by higher $N$. The curves converge rapidly within a few iterations, demonstrating the efficiency of the proposed optimization algorithm. It is also worth noting that the utility function reaches a performance ceiling that reflects the influence of the transmit power budget $P_{\rm \max}$.

	\begin{figure}[h!]
		\centering
		 \vspace*{-0.05in}
         \hspace*{-0.1in}
		 \includegraphics[height=3in,width=3.75in]{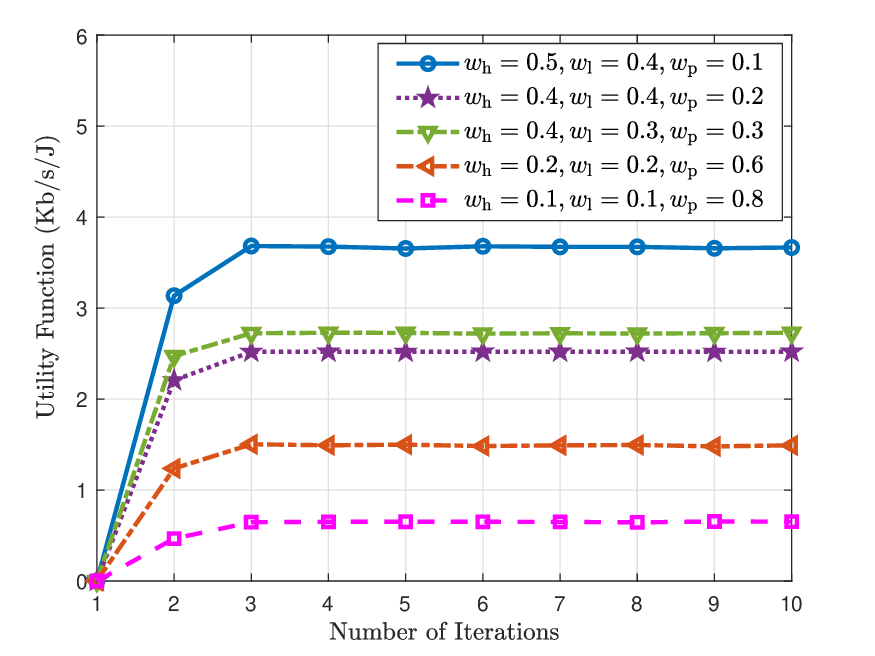}
		\caption{Utility function versus the number of iterations. Here, $P_{\rm \max}=50$ dB.}	
		\label{fig14}
	\end{figure}

    Figure~\ref{fig14} shows the impact of different weight combinations on the utility function of the DBRIS-aided SWIPT system. Prioritizing data rates while assigning a low weight to the total power consumption of the system, e.g., $w_{\rm h} = 0.5$, $w_{\rm l} = 0.4$, $w_{\rm p} = 0.1$, yields the highest utility. Conversely, emphasizing a higher weight factor for the total power consumption,
    e.g., \( w_{\rm p} = 0.8 \), degrades the performance of the SWIPT system under study. Here, all configurations with different weights converge within a few iterations, highlighting the efficiency of the optimization process and the role of weight selection in achieving the desired performance trade-offs.
	\section{Conclusion}	
	This paper introduced the integration of DBRIS in wireless system design, and investigated the energy efficiency gains in two representative systems, namely, DBRIS-assisted communication and DBRIS-aided SWIPT. For both use cases, the energy efficiency maximization problem was formulated and solved using alternating optimization techniques for the joint phase-shift design and power allocation across different frequency bands. Numerical results demonstrated the viability and effectiveness of the DBRIS in enhancing the system performance compared to operations with multiple SBRISs. These findings align with the potential of the DBRIS technology in addressing challenges such as hardware redundancy and spatial limitations, and pave the way for effective solutions towards improving the sustainability of next-generation wireless systems as well as upgrading existing ones. Future directions of this research work include extending the proposed DBRIS-aided communication and SWIPT models to multi-user scenarios, quantifying the gains of the DBRIS integration in other uses cases, modelling the mutual coupling using a testbed, investigating the impact of hardware impairment on system performance, etc.
	
	\bibliographystyle{IEEEtran}
	\bibliography{references}
\end{document}